UC BERKELEY
CENTER FOR LONG-TERM CYBERSECURITY

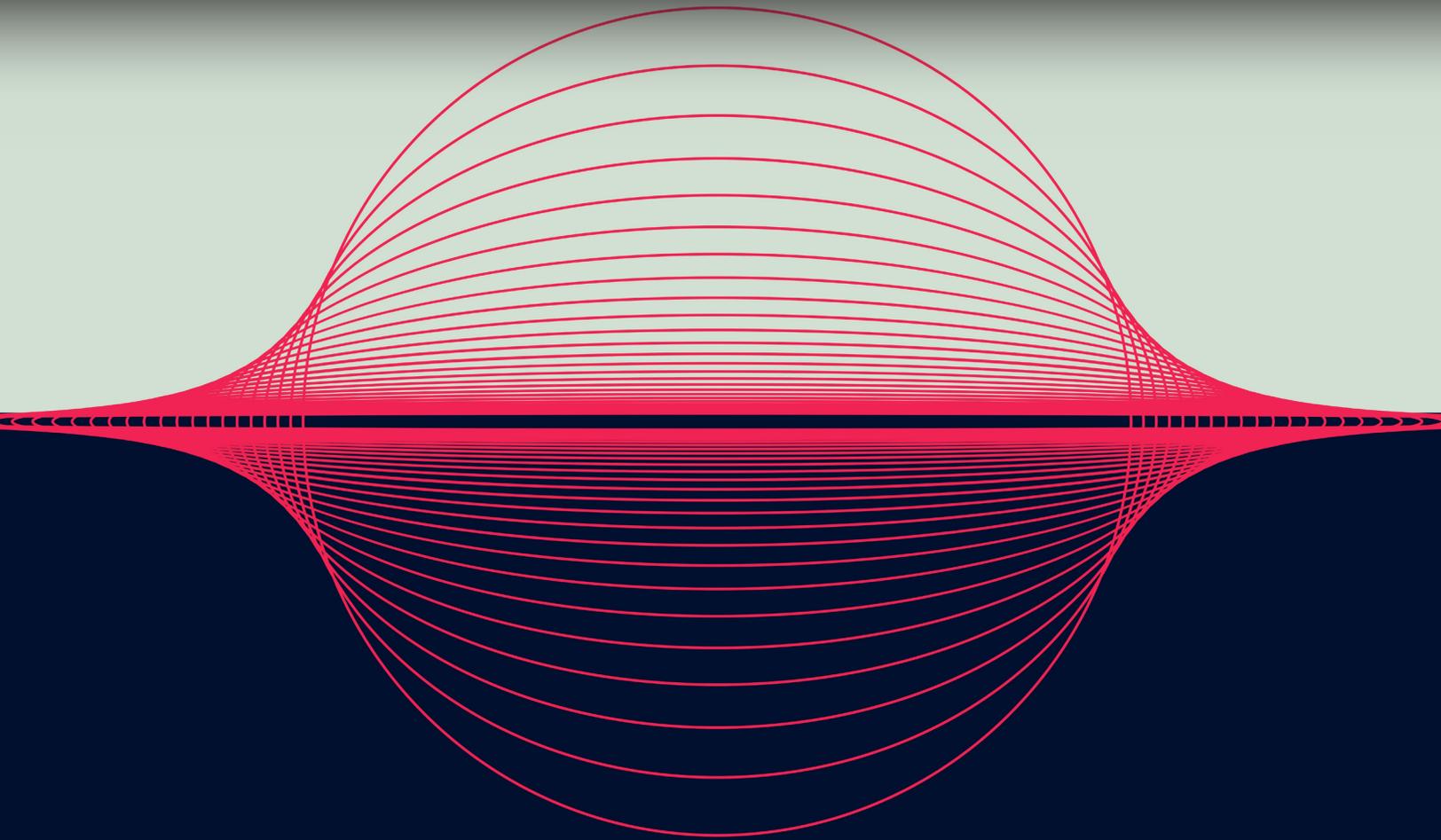

# Intolerable Risk Threshold Recommendations for Artificial Intelligence

**KEY PRINCIPLES, CONSIDERATIONS, AND CASE STUDIES TO INFORM FRONTIER AI SAFETY FRAMEWORKS FOR INDUSTRY AND GOVERNMENT**

DEEPIKA RAMAN | NADA MADKOUR | EVAN R. MURPHY
KRYSTAL JACKSON | JESSICA NEWMAN

February 2025

# Intolerable Risk Threshold Recommendations for Artificial Intelligence

**Key Principles, Considerations, and Case Studies to Inform Frontier AI Safety Frameworks for Industry and Government**

Deepika Raman, Nada Madkour, Evan R. Murphy, Krystal Jackson, Jessica Newman

UC Berkeley Center for Long-Term Cybersecurity

# Table of Contents

# Executive Summary

With increasing capabilities and widespread deployment, frontier AI models — highly capable foundation models on the cutting edge of artificial intelligence technology — may pose intolerable risks due to their misuse, malfunction, or systemic factors. For example, frontier models could lower barriers for a terrorist, state-affiliated threat actor, or other adversary seeking to cause a high-impact events, such as a chemical, biological, radiological, or nuclear (CBRN) attack, or could result in illegal discrimination against large numbers of people when adopted for downstream applications without sufficient evaluation.

Given the potential harms of foundation models, AI developers need to establish limits, or "thresholds," to ensure that AI-based technologies are designed to prevent severe harms, or "intolerable risks," from becoming reality. This paper provides an overview of intolerable risk thresholds for AI, and proposes recommendations for organizations and governments exploring how to define and implement these thresholds. We define intolerable risks as **risks of severe harm to public safety and human rights, inequality, economic loss, or an unwelcome alteration of values and societal norms that can manifest through purposeful direct adversarial misuse, systems failures, unintended consequences arising from a chain of related events, or cascading, secondary, or simultaneous failures of frontier models.**

In our discussion, we include model harm arising from the risk categories of CBRN weapons, cyber attacks, model autonomy, persuasion and manipulation, deception, toxicity, discrimination and socioeconomic disruption. Appendices A and B offer detailed background on the various risk taxonomies and risk actors that we considered from current AI governance literature to help define our scope.

A survey of the various techniques used in AI evaluations, and a review of the initial efforts across industry, academia, AI Safety Institutes (AISIs), and regulators to establish thresholds, showed that evaluations of AI systems to set capability thresholds, do not necessarily replace estimations of risk, and instead can encode risk tolerances in more implicit ways (Campos et al. 2024). Through this paper, we provide a number of key principles and considerations to help quantify some of these impacts through a discussion on adapting risk assessment frameworks from other industries. We also propose a number of considerations for organizations and governments exploring how to define and operationalize intolerable risk thresholds in Section 3:

- Design thresholds with adequate margins of safety to accommodate uncertainties in risk estimation and mitigation.
- Evaluate dual-use capabilities and other capability metrics, capability interactions, and model interactions through benchmarks, red team evaluations, and other best practices.
- Identify "minimal" and "substantial" increases in risk by comparing to appropriate base cases.
- Quantify the impact and likelihood of risks by identifying the types of harms and modeling the severity of their impacts.
- Supplement risk estimation exercises with qualitative approaches to impact assessment.

- Calibrate uncertainties and identify intolerable levels of risk by mapping the likelihood of intolerable outcomes to the potential levels of severity.
- Establish thresholds through multi-stakeholder deliberations and incentivize compliance through an affirmative safety approach.

Based on these principles, Section 4 presents clear threshold recommendations for some intolerable outcomes in the identified risk categories. We conclude our discussion with three case studies that detail the unique ways in which the key principles (see Section 3) can be applied to further operationalize our threshold recommendations (see Section 4) for three distinct types of intolerable risks: misuse risks from CBRN capabilities, cross-cutting risks from evaluation deception, and systemic risks from AI-generated misinformation.

This paper was supported by several rounds of stakeholder consultations, including roundtable discussions, workshops, and feedback from various experts. We intend for this material to be a starting point or supplementary resource for others to use in their own deliberations to begin timely efforts to implement policies that prevent intolerable risks from ever occurring (ex ante), rather than merely implementing safeguards in response to their occurrence (ex post). Through this work, we hope to advocate for **"good, not perfect"** thresholds and to err on the side of safety in the face of uncertainty and limited available data.

# Acknowledgments

This working paper was informed by an in-person roundtable held in Berkeley, California on November 12, 2024, and a virtual workshop held on December 16, 2024, that brought together representatives from academia, industry, civil society, and government. We are grateful to the participants for their expert feedback and insights. Participants do not necessarily endorse the recommendations or considerations included here. Participants in the November and/or December workshops included: Sacha Alanoca, Esteban Arcaute, Anthony Barrett, Steph Batalis, Colleen Chien, William Cutler, Tarana Damania, Ian Eisenberg, Florence G'sell, Giulia Geneletti, Paolo Giudici, Jason Green-Lowe, David Evan Harris, Yolanda Lannquist, Owen Larter, Toni Lorente, Richard Mallah, Mina Narayanan, Cassidy Nelson, Joe O'Brien, Neev Parikh, Daniel Schiff, Colin Shea-Blymyer, Olivia Shoemaker, Risto Uuk, Meredith Lee, Alisar Mustafa, Chris Meserole, Stuart Russell, Maeve Ryan, Madhu Srikumar, Shane Witnov, and others. We are grateful to the Foundational Research Grants program at the Center for Security and Emerging Technology (CSET) for support of our ongoing work in this space.

# 1. Introduction

## 1.1. The Call for Defining Intolerable Risk Thresholds

In May 2024 at the AI Seoul Summit, 16 global AI technology companies committed to publishing their efforts to measure and manage risks posed by their frontier AI models in an accountable and transparent manner, and committed to determining thresholds for intolerable risks. Specifically, as part of these **Frontier AI Safety Commitments**, the organizations are required to define **"thresholds at which severe risks posed by a model or system, unless adequately mitigated, would be deemed intolerable"** (DSIT 2024a).

At the same summit in Seoul, **27 nations and the EU announced their intent to define these thresholds for frontier AI systems in advance of the AI Action Summit in France,** signaling the regulatory appetite to challenge self-imposed, "voluntary" limits set by industry actors (DSIT 2024b). Both pledges explicitly state the importance of ensuring these thresholds are defined with input from a range of trusted actors, and require reporting on the different capacities of their involvement in these efforts.

While the idea of "intolerable risks" from frontier models might seem futuristic, we are already seeing severe impacts from current model deployments — for instance, from AI-generated misinformation that is spreading widely across our information ecosystems and undermining social trust. While (arguably) not yet at an intolerable level, the growing capabilities and wide-scale deployment of these models should serve as a warning to undertake immediate tangible efforts to mitigate such risks.

In its most substantial articulation of national security strategy and policy toward AI, the Biden Administration released the National Security Memorandum (NSM) on artificial intelligence, which was issued in October 2024 and called for federal agencies to identify and prohibit "unacceptable levels of risk" from AI applications (White House 2024a).[1] Accompanying this memo was the Framework to Advance AI Governance and Risk Management in National Security, which further detailed a principled guide to determine the threshold of unacceptable risk (White House 2024b). Previously, the National Institute of Standards and Technology (NIST) released specific guidance to manage intolerable risks arising from the misuse of frontier models (NIST 2024a). The National Telecommunications and Information Administration (NTIA) has called for the federal government to maintain a portfolio of risk cases and thresholds in the context of dual-use, open, foundational models, calling for special attention to intolerable risks to aid effective monitoring and management (NTIA 2024).

Several frameworks for pre-release risk assessment and decision-making include dual-use capability evaluation and some form of explicit or implicit thresholds for dual-use capability hazards that should be regarded as intolerable (Anthropic 2024, Google DeepMind 2024, OpenAI 2023b). The Organization for

---

[1] The majority of this paper was drafted and finalized prior to the rescission of Executive Order 14110 on January 20, 2025.

Economic Cooperation and Development (OECD) and the Safe AI Forum (SAIF) among others have also launched working groups and requested public inputs to determine early warnings and risk thresholds to safeguard frontier AI systems (OECD 2024, IDAIS 2024). One such consultation report (TFS 2024) from the Future Society and its partners, released in January 2025, reiterates the need for stakeholders at the upcoming AI Action Summit in Paris to supplement the voluntary commitments from industry with concrete thresholds. **This report calls on the Summit's participants and the network of AI Safety Institutes to provide a timely solution and present a concrete roadmap for model developers and evaluators to operationalize clear thresholds to mitigate intolerable AI risks.**

## 1.2. Purpose and Scope

In this paper, we provide a background on intolerable risk thresholds and propose a number of recommendations and considerations for developers, third-party organizations, and governments, that are exploring how to define and operationalize intolerable risk thresholds. We intend for this material to be a starting point or supplementary resource for others to use in their own deliberations.

In Section 2, we aim to provide a definition for intolerable risks through a discussion of different policy approaches, risk taxonomies, and AI safety frameworks. We do not recommend an exhaustive risk taxonomy,[2] standardized severity scales, or definitive likelihood estimations. Instead we collate past precedent and current practices on these measures, and provide our recommendations to effectively combine these approaches through the coordination of multiple cross-disciplinary stakeholders. While trends in AI development, risk mitigation, and management techniques[3] were explored during the drafting of this paper, these topics are not directly addressed in this paper.

We offer specific guidance and recommendations on setting quantitative and qualitative thresholds (see Section 3), but we do not explicitly draw red lines for intolerable outcomes. Our work provides a recommended template for policymakers, model developers, and model providers to utilize when operationalizing risk thresholds. However, our primary focus is on how risk thresholds can be operationalized by upstream AI developers or codified by regulators (rather than downstream AI "systems").[4] In Section 5, we provide case studies to further illustrate the operationalization of the recommendations from Section 3.

## 1.3. Evolution of this paper

Following an in-person roundtable discussion held on November 12, 2024 at the University of California, Berkeley campus, a first draft of this paper was published in November 2024 to invite stakeholder feedback (Barrett et al. 2024a). Following this release, a virtual workshop was held on December 16, 2024 to draw feedback, which greatly influenced the revision of this paper, along with feedback sent to us via email.

---

[2] See Appendix A for a discussion on why this needs to be a subjective approach
[3] For more on AI risk mitigation and management, see Barrett et al. (2025).
[4] Our focus on AI models, rather than AI systems, is intended to reflect current trends in AI development and governance, in which a single large GPAI/foundation model typically plays a central role as a core part of either a GPAIS or a relatively narrow-purpose end use application. Our focus includes but is not limited to general purpose AI (GPAI), Generative AI (GAI), and agentic AI.

The working paper released in November 2024 defined capability-based categories for intolerable malicious use risks, but did not include other unacceptable uses, impacts, and limitations under the "intolerable" umbrella. However, as has become evident from recent policy developments (e.g., the second draft of the General-Purpose AI Code of Practice, EC 2024[5]), and echoed by our workshops, the highest probability of insidious risks from AI deployments will likely be the cumulative effect of "high-impact" AI systems that operate in critical domains. As these systems are increasingly adopted for applications in such domains, they could lead to long-term effects such as unmitigated workplace automation or large scale algorithmic discrimination causing widespread economic losses that exacerbate existing inequalities and power imbalances, etc. Managing these risks will require commitments from all actors in the AI ecosystem, keen monitoring and oversight, and resistance to an inherently tech-first approach to designing access to critical infrastructure and services.

The main changes between the working paper from November 2024, and this version include the extension of intolerable risks beyond capability-based risks, updated considerations to help broaden the focus of intolerable risk thresholds beyond capability evaluations, and the addition of illustrative case studies to demonstrate the operational feasibility of our recommendations.

---

[5] The second draft of the EU General-Purpose AI Code of Practice was published on December 19, 2024. The final version is planned for publication in May 2025. (See, EC 2025 for more in the EU Code of Practice timeline.)

# 2. Background Analysis on Intolerable Risks and Thresholds

As recounted in the introduction, the Frontier AI Safety Commitments seek to define "thresholds at which severe risks posed by a model or system, unless adequately mitigated, would be deemed intolerable" (DSIT 2024a).

**Expanding on Advanced Capabilities-Based Risks**

Current industry frameworks on intolerable risks from frontier models often highlight risks that arise from catastrophic events (e.g., deployment of bioweapons) arising from model capabilities and their misuse. Across these frameworks, there is relative consensus on the key risk categories in scope. The frameworks from Google, OpenAI, and Anthropic discuss model capabilities, such as ability to develop CBRN weapons, conduct cyber attacks, and autonomous capabilities, with a subset of frameworks additionally considering models' capacity for persuasion and to some extent, deception as other risk categories to potentially track (Anthropic 2024, Google DeepMind 2024, OpenAI 2023b).

By continuing to narrow our focus to these risks from misuse of these advanced (at times futuristic) capabilities, we risk over-indexing on specialized thresholds at the expense of tackling current AI hazards (Khlaaf et al. 2024). This narrow scope sidelines emerging risks from current model capabilities and limitations (Raji et al. 2022), especially in combination with smaller, domain specialized AI tools that may pose significant harms. The capability-based categorization also leaves out other systemic risks (EP 2024) like the long-term impacts of frontier models (e.g., deterioration of democratic norms, large-scale discrimination) that could fundamentally change the fabric of society, sidelining industry's accountability for such outcomes (Critch & Russell 2023). This paper is a reflection of the evolving considerations for the different categories of intolerable risks to be considered (as demonstrated by the changes in our research scope, traced in Section 1.3).

Without a common definition, scale, or policy to decisively determine the gamut of "intolerable risks," we begin with an incomplete premise on which to base our threshold-setting exercise. Therefore, by reflecting on the various risk taxonomies, motivations of risk actors, and objects at risk, we define the scope of intolerable risks as emanating not only from singular events with potentially catastrophic impacts that could threaten national security, but also from large-scale societal risks resulting from the emergent behaviors, vulnerabilities, and capabilities of AI models.

**Defining Intolerable Risks**

Based on our literature review and scoping exercise, **we define "intolerable AI risks" as risks of severe harm to public safety, human rights, inequality, economic loss, or an unwelcome alteration of values and societal norms that can manifest through purposeful adversarial** **misuse, systems failures, unintended consequences arising from a chain of related events, or cascading, secondary, or simultaneous failures of frontier models.**

## 2.1. Risks in Scope

Previous work on risk taxonomies that aims to classify potential harms primarily analyzes the scale of AI adoption or the societal impact of AI models along the lines of bias, misinformation, automation, socioeconomic and environmental harms, etc. (Shelby et al. 2023; Vidgen et al. 2024; Weidinger et al. 2022). In measuring these risks, there is growing consensus on not only identifying the capabilities of frontier models but also designing how such models are evaluated in the context of their interaction with society (Solaiman et al. 2024). Additional categorizations of risk that inform our framing of intolerable risks include– capabilities exploitable by malicious actors, risks stemming from product or model malfunction, systemic risks, and other cross-cutting risk factors (Autio et al. 2024, Bengio et al. 2024, 2025).

To limit the scope of our study on intolerable risks, we chose the following (non-exhaustive) set of risks (from model misuse, malfunction, and their compounding effects) in line with our definition of intolerable risks.[6] These categories also feature in most policies from AI developers and government actors as priority areas for monitoring and risk mitigation. These risks are not mutually exclusive: a bad actor could use a model with a high degree of autonomy to aid in a cyber attack, for example. We outline further rationale for these risks' inclusion, and the potential intolerable outcomes they pose, in Section 4, Table 1.

**Selected Risk Categories:**

- Chemical, Biological, Radiological, and Nuclear (CBRN) Weapons
- Cyber Attacks
- Model Autonomy (loss of human oversight)
- Persuasion and Manipulation
- Deception
- Toxicity (including CSAM, NCII)
- Discrimination
- Socioeconomic Disruption

Other crucial ways to identify risks beyond those explored in existing literature is through extensive open-ended red reaming activities, as well as other risk estimation techniques (Campos et al. 2024), which are discussed in Section 2.3.

---

[6] While this paper does not currently include the environmental risks from AI in its scope, it cannot be dismissed that rapidly improving machine learning capabilities, while beneficial for many applications, may also increase the risk of short-term economic incentives sidelining long-term sustainability goals. Tracking energy, water usage, compute, runtime, and carbon emissions or holistic life cycle assessment (Berthelot 2024) can help determine environmental impacts of AI. Designation of such risks as 'intolerable' should be supplemented by evaluating the common argument of risk-benefit trade offs.

## 2.2. Intolerable Risk Thresholds

The Frontier AI Safety Commitments (DSIT 2024a) characterize intolerable risk thresholds as follows:

> "Thresholds can be defined using model capabilities, estimates of risk, implemented safeguards, deployment contexts and/or other relevant risk factors. It should be possible to assess whether thresholds have been breached."

The second draft EU AI Code of Practice, similarly advocates for the articulation of "…conditions under which further development and deployment of a general-purpose AI model with systemic risk will not proceed due to insufficient mitigations for (a) keeping risk below an unacceptable level, or (b) appropriately mitigating risk that is below an unacceptable level" (EC 2024, p52).

Commonly recommended mechanisms to assess and manage intolerable risks include **capability thresholds, compute thresholds, and risk thresholds.** These are discussed further in Appendix B.

While many factors contribute to a model's risks and capabilities, compute (i.e., the computational resources required for an AI model) has emerged as the common initial metric to identify models that require further regulatory oversight and evaluation. Both the former White House Executive Order 14110[7] (White House 2023) and the EU AI Act (EP 2024) have made use of compute thresholds to categorize high-risk models.

Koessler et al. (2024) have argued for defining thresholds primarily in terms of "risk" (i.e., likelihood and impact). The literature on risk management for rare or novel catastrophic events shows how risk estimations for such events introduce debatable assumptions or must factor in great uncertainties. In the absence of regulatory guidance, such estimates may not help draw a clear threshold "line" but rather suggest a very broad plausible range.

Model capability often serves as an (imperfect) proxy for risk. With no standardized methods to measure compute power, the dual-use nature of foundation models, the current narrow focus on misuse risks, and the lack of reliable risk estimates, current industry frameworks prominently feature capability thresholds that are most closely aligned with establishing thresholds for intolerable risk. These metrics do not carry the compounding uncertainties of likelihood estimates, but still take impact into account. Appendix A details the various risk levels that model developers and deployers have identified to track organizational capability thresholds.

Evaluating the current approaches against the mandate of the Frontier Safety Commitments makes it evident that all three types of thresholds differ widely in the reliability of their estimation of risk, as well as the feasibility of their operationalization. It is also evident that, despite the current focus on capability thresholds in frontier frameworks, these evaluations do not necessarily replace estimations of risk, and instead encode risk tolerances in more implicit ways (Campos et al. 2024). In the absence of the perfect method, **we recommend that multiple evaluation criteria be considered in tandem when singular measures are hard to quantify**, in order to reliably establish thresholds to prevent intolerable risks.

---

[7] The majority of this paper was drafted and finalized prior to the rescission of Executive Order 14110 on January 20, 2025.

> Given the devastating potential of intolerable AI risks, it is imperative to implement policies to prevent the harms from ever occurring (ex ante) rather than merely implementing safeguards in response to their occurrence (ex post). When developing risk thresholds in this context, empirical research is a highly scarce resource, and it is important to strive for **"good, not perfect"** thresholds and to err on the side of safety in the face of uncertainty and limited available data.

Based on this motivation, this paper recommends establishing intolerable risk thresholds that reflect the complex impacts that emerge from model capabilities, deployment contexts, and the likelihood of different outcomes (for more on this, refer to Clymer et al. 2024, Section 6.6). Thresholds must also be designed to account for diverse impacts, such as loss of human lives, decline in quality of life, threats to human rights, damage to property and infrastructure, financial losses, environmental impacts, etc.. They should also appropriately address the potential for compounding harms from small-scale or less severe risks, in addition to mitigating risks from singular catastrophic events.

While risk thresholds are not commonly operationalized for AI risks, we hope to provide some guidance and recommendations through this paper to help accelerate their adoption for AI governance.

## 2.3. Risk Estimation

There are a variety of **negative outcomes** that could come about from the risks listed above (Section 2.1). The outcomes we are most concerned with preventing are systemic, widespread, or have far-reaching consequences with large-scale societal impact. In simple terms, **risk can be defined in terms of likelihood (probability of an event or negative outcome), and severity of harm (magnitude of impact).**

### Discussion on AI Risk Estimation

A quantitative measurement of risk can be extremely valuable, but may not always be feasible, especially for low-probability, high-impact events that are harder to estimate. Despite the difficulty in reliably estimating risks, or precisely because of its challenging nature, we believe that it is doubly important that risk estimations of singular events that could cause catastrophic impacts remain a responsibility of the frontier model developers, and not downstream deployers.

Since intolerable risks go beyond singular catastrophic events and also need to account for the long-term compounding nature of societal-scale risks, **there is a need for researchers and civil society to evaluate the severity of systemic risks, for government regulators to explicitly establish acceptable harm severity levels for society, and for developers to ensure that models do not** **cross these levels** in order to enforce better governance of model development. This is not to say that catastrophic risks from singular incidents do not need state intervention; operationalization of all threshold risks can only be reasonably and reliably achieved through the collaboration of state and industry efforts.

Koessler et al. (2024) identify the different risk measurement approaches from industry frameworks– risk models, threat models, and risk scenarios (Anthropic 2024, OpenAI 2023b, Google DeepMind 2024) — to demonstrate the pathways between risk factors and harmful impacts. We similarly recommend the use of risk scenarios for establishing intolerable outcomes, to help create a robust understanding of model capabilities, the likelihood of such risks occurring from model use and deployment, and their impacts should they manifest. Harms can be categorized based on who or what is impacted, and severity levels assigned can be based on the magnitude of impact. We present additional recommendations from emerging risk estimation frameworks and other industries in the rest of this section.

Comprehensive risk models containing all possible risk scenarios are extremely difficult to develop, and it is recommended to start with a limited and defined number of risk scenarios (Koessler et al. 2024). To the greatest extent possible, regulators and government actors — in partnership with academia, civil society, industry, and impacted communities — should create an exhaustive list of risks and their connection to negative outcomes, then prioritize both risks and outcomes using a combination of likelihood and impact.

Regardless of the risk measurement method, **it is important that risk thresholds are operationalized and specific enough to ensure that multiple evaluators with access to the same resources would agree on the risk threshold determination of an evaluated model** (DSIT 2023a).

Despite the rich literature that tracks the identification, assessment, and management of AI-related risks, there is no clear consensus on a favorable methodology to assess the likelihoods of different intolerable outcomes, or the resulting severity of their harms. Estimating the likelihood of risks is especially important given the increase in systemic risks with not just model capabilities but also model reach (EP 2024, Recital 110).[8] Therefore, we strongly recommend that determining thresholds require the involvement of a diverse set of stakeholders with expertise in areas that extend beyond frontier AI, or even the risk area itself, such as national security, environmental science, human rights, etc. Apart from multi-stakeholder participation, looking to other high-risk industries that have more mature risk management strategies could also provide important lessons for AI risk estimation.

### Precedents from Other Industries

We have a wealth of examples of risk assessment strategies from fields such as aviation, healthcare, nuclear energy, and chemical manufacturing that detail frameworks to map and safely monitor the impacts and probabilities of industry-specific risks (Tudoran 2018, Dezfuli et al. 2014). The FDA, for instance, extrapolates the likelihood of device malfunction, likelihood of harm to patient, and the total number of patients exposed to inform evaluations of safety compliance from medical device manufacturers during clinical trials (FDA 2016). Precedents from other industries show unique ways of identifying risk categories and calculating acceptable thresholds, and there is no one-size-fits-all approach. While methods to characterize different harms and calibrate the appropriate scales of impact are common to most risk assessment methodologies, the approaches used are diverse. Some risks are tightly coupled with harms (e.g., a chemical spill causing negative health outcomes), while AI risks could be tightly connected to capabilities or knowledge domains. However, other sectors have grappled with accounting for nuance in the chain of harm. For instance, to prevent misuse, the FAA prioritizes identifying hazards around access control, personnel screening, and protection against unauthorized activities.

---

[8] And in the absence of consensus on likelihood estimations, introducing subjective probabilities could help adequately capture the uncertainty inherent to such risk modeling. (Flage et al. 2014).

### Adapted to AI Risk Estimation

The literature on catastrophic AI risks surfaces several approaches that adapt industry precedents and modeling paradigms to assess risks from AI. For instance, fault tree analysis and event tree analysis are commonly used to determine catastrophic AI risks (Barrett & Baum, 2017a, 2017b). For a detailed review of industry precedents adapted to assess AI-specific risks, see Koessler and Schuett (2023). Recent work on AI risk has expanded to encompass a wider variety of risks, some of which are illustrated below.

- One approach derives **harm severity levels** by classifying observed AI incidents into risk types, then using the CSET AI Harm Framework to categorize tangible and intangible harms (Hoffman and Frase 2023). This framework generally uses a logarithmic scale to emphasize the rapidly escalating nature of potential damage that could be caused by AI models. This type of framework could be useful in setting uniform standards for intolerable outcomes across multiple risk domains.

- Another approach to consider is the **Probabilistic Risk Assessment** (PRA) framework, a risk matrix that is commonly used across industries, for instance, its application to the healthcare industry can be seen in Pascarella et al. (2021). By adapting this popular technique to assess AI risks, the PRA workbook for AI risks designed by the Center for AI Risk Management & Alignment helps surface potential future threats, or "known and unknown unknowns," which are important considerations when modeling emerging AI risks. This approach examines risks linked to AI aspect groups, such as capabilities, knowledge domains, and sociotechnical impact domains. This analysis examines both direct and indirect risks to individuals, society, and the biosphere. Walking through the methodology provides assessors with a report card detailing risk areas and levels. The PRA models AI harms based on potential to rapidly escalate and cascade (Wisakanto et al. 2025).

These approaches seem promising but have not yet been widely or uniformly applied in estimating AI risks. Nonetheless, we suspect that frameworks like harm severity scales and probabilistic risk assessments will be widely adopted tools that can help calibrate the uncertainties inherent to risk estimation through standardized scales. That said, we are not yet confident in recommending any singular framework as a standalone assessment that regulators and developers should rely on. We instead distill such efforts and use illustrative steps in the following section to operationalize risk thresholds.

# 3. Key Principles and Considerations

Determining risk thresholds for dual-use foundational models is an exercise in balancing overarching thresholds with wide applicability and domain-specific thresholds with high specificity. This specificity may reflect the range of actors, norms, practices, and technical systems involved, and the specific ways in which risks may emerge in that domain (Shelby et al. 2023). As illustrated in the section above, these efforts require the involvement of governments and innovators to develop and operationalize thresholds that reflect a range of aspects across the lifecycle of frontier model development, as well as model deployment (NTIA 2024).

Additionally, we see different risk categories applying the key considerations illustrated in this section in unique combinations to determine thresholds and corresponding actions, as illustrated through the case studies in Section 5. For instance, if strong correlating metrics or uplift studies can be designed for capability-based risks like CBRN weapons, strict capability thresholds could be established to stall model development or deployment. On the other hand, negative outcomes from model limitations, or compounding risks from their deployment context (for example, misinformation risks), may require evaluating risk estimates and model propensities in tandem. This approach may lead to efforts to curtail deployments in certain application areas, or mandate that mitigation or alignment efforts accompany any substantial increases, rather than an altogether halting of development.

This paper builds on frontier model safety frameworks and prominent policy language in identifying some intolerable risks in scope (Section 2.1). But as we note, this is not an exhaustive list, and policy frameworks should amalgamate different risks and apply their societal and cultural perspectives toward defining intolerable risk categories. (For further discussion on these subjective estimations of risk, see Appendix A.) There are a range of taxonomies that can be mapped to appropriate frameworks for interoperability, and they must be standardized where feasible to enable easier adoption in the risk assessment and reporting exercise. This is an area where government agencies, policies, and international agreements can lend their efforts to operationalize better oversight.

The recommendations in Section 3.1 are intended for numerous actors in the AI development and deployment lifecycle, including but not limited to, academic institutions, industry, and government actors. Section 3.2, Codifying Thresholds into Regulations, is mainly intended for government actors.

## 3.1. Foundations for Establishing and Operationalizing Thresholds

**Overarching Guidance**

- **Account for Uncertainty:**

    - Standardized scales, such as the harm severity levels or PRA framework discussed in Section 2.3, can serve as critical approaches to calibrate uncertainty across the various types of harms caused by AI risks. Additionally, depending on the statistical approach and risk assessment design, AI developers may aim to evaluate whether 1) the mean value of uplift is less than the threshold, or 2) the upper end of the confidence interval for the value of uplift is less than the threshold.[9]

    - **Purpose Limitation:** When uncertainty in risk estimation can interfere with decision-making for model deployment in high-impact application areas, like healthcare or financial lending, cautionary governance must be practiced. Similar to the prohibited AI practices from the EU AI Act (EP 2024, Article 5), or prohibited AI use cases for government agencies from the U.S. White House Framework to Advance AI Governance and Risk Management in National Security (White House 2024b), regulators can disallow the deployment of AI models in certain use cases where risk estimations are accompanied by high levels of uncertainty, before critical risk levels are reached. This can be operationalized especially effectively when taking a sector-specific approach in establishing critical and intolerable levels[10] of risk.

- **Leave some Margin of Safety**
  Because of the uncertainties of estimating risk, limitations in eliciting capabilities, and the growing category and scale of AI risks, it is important that thresholds be set at highly conservative levels, but designed with adequate flexibility so that they can evolve in the light of rigorous assessments and robust mitigation.

    - **To account for open source models:** Design thresholds that take into consideration the unreliability of virtually all safeguards and risk mitigation efforts at this time. Anthropic defines redline CBRN capabilities in terms of reaching human level expertise. However, it is

---

[9] A confidence interval overlapping with the intolerable risk threshold should trigger the need for more work with a larger sample size to reduce the uncertainty range and give a better assessment of whether the effect is actually below the intolerable risk threshold.

[10] We define critical levels of risk as determined by any "substantial" increases in capability or risk. This critical level of risk is distinctively lower than an intolerable level of risk. For example, models with even the lowest level of risk for privacy breaches can be unsuitable in processing sensitive health or voter information, but may be suitable for other sectors. See Section 5b, "Determining Which Artificial Intelligence Is Presumed to Be Safety-Impacting or Rights Impacting," in OMB (2024).

necessary to leave a margin of safety, especially for models intended for open weights release.[11] it is worth aiming to stay well below human expert level thresholds (e.g., setting a threshold at halfway to human expert levels). More generally, it seems prudent to operationalize intolerable-risk thresholds at approximately the "substantial" level, leaving some margin of safety before arriving at a "severe" level.

- **To account for limitations in mitigations:** Some intolerable risk thresholds should not factor in model guardrails or other model-capability mitigation measures. Virtually all guardrails for capability are inadequate or unreliable and can be trivially circumvented via jailbreaks (El-Mhamdi et al. 2022, Wu et al. 2024) or reversed via fine tuning(e.g., (Carlini et al. 2023), at least at this point in time (Zou, Wang et al. 2023).

- **Future-Proofing Thresholds:** Intolerable risks do not necessarily require large-scale runs. Therefore, **due consideration needs to be placed on how thresholds might have to change** rapidly with the widespread availability and affordability of compute power for fine-tuning open models (Seger et al. 2023). Entrench these margins of safety in threshold determination along the dimensions of **increasing affordability, access, and expertise in AI models** to ensure sufficient safety between calibrations.

- **Transparent Reporting**

  - Documented risks and decisions should also be reported transparently — to regulators or internal review boards, red teamers, and auditors — to ensure appropriate testing against vulnerabilities in the chosen design of the model. Additionally, limitations and uncertainty should also be documented and reported for all safety evaluations.

## Capability Evaluations

- **Capability Evaluation Methods**

  At least two methods are available to evaluate model capability: open benchmarks and closed red teams. **Benchmarks** utilize a standardized set of questions and answers through model prompts to evaluate model capability, making them a quick and cost-effective option.[12] **Red team evaluations** involve intensive and interactive testing by domain experts, and achieve a higher level of accuracy by incorporating sensitive details. The in-depth and labor-intensive nature of red-team evaluations render them higher in cost. Barrett et al. (2024b) recommend utilizing open benchmarks as a quick, cost-effective preliminary evaluation of model capabilities, and running in-depth red team evaluations if a model receives a benchmark score indicative of high dual-use potential.

---

[11] Models with an open weight release tend to be easiest to fine-tune or enhance in other ways (e.g., reinforcement learning and chain of thought to add capabilities), but they cannot be monitored or decommissioned by the model developer through an API.

[12] An example of a CBRN and cyber-related benchmark is WMDP (Li et al. 2024a,b,c). Other capability benchmarks include PlanBench (Valmeekam 2022a,b), WorldSense (Benchekroun et al. 2023a,b), and MMLU (Hendrycks et al. 2020). While quick and cost-effective, benchmarks lack accuracy. For other socio technical safety evaluations, see [this repository](#).

- **Compare to Appropriate Base Cases**

    Human-uplift studies can be helpful in calculating capability-enabled risks by comparing AI-augmented risk scenarios to the baseline performance and likelihood associated with such intolerable outcomes sans AI assistance. Such studies enable quantitative risk assessments by mapping potential technological enhancements against current performance thresholds, which allows precise calculation of scenarios where capability gains might introduce intolerable risks. This can be used to further inform model capability assessments and the establishment of capability-based thresholds. Typically with human-uplift studies that evaluate CBRN and cyber risks, dual-use capability assessment methods either implicitly or explicitly compare a model's outputs to information available from internet searches. (See more in the CBRN weapons case study in Section 5 below, or Mouton et al. 2024, Patwardhan et al. 2024, or Dubey et al. 2024).

    - For assessing marginal risks of releasing a particular model, it could be valuable to compare a new LLM to other available LLMs, instead of to the Web. However, a model's outputs should not only be compared to other available models. Closed-weights models can be rolled back, they can be made unavailable very quickly via the provider's control of an API, but open-weights models cannot effectively be made unavailable after release of their weights. With growing investment in AI globally and an increasing number of models released each year, using existing models as a baseline could easily lead to an exponential growth in risk from AI models overall. In particular, open-weights model releases, cannot be contingent upon comparisons between the rising marginal-risk of their current and previous open-weights release. That would be a slippery slope, and bad risk management policy.

- **Identify Substantial Increases in Risk**

    At a minimum, aim to identify cases of substantial increase to marginal risk. The concept of marginal risk[13] can be a useful way to compare the risks of a model compared to standard tools, such as searches on the internet. We can assume several terms as approximately equivalent to "substantial," but without getting into a legal analysis of such terms, we define "substantial" here to mean something greater than detectable.

---

[13] The concept of marginal risk can also contribute to a slippery slope, as progressively worse and more dangerous models are considered acceptable. It is not appropriate to compare the risks of a new frontier model to the risks of every other available model because there are already widely proliferated models without sufficient safeguards that can be used to cause significant harm. (This is also recounted in the subsection on "Compare to Appropriate Base Cases," above).

- **Capability Metrics to Evaluate Impact**

    Several aspects of **model capabilities must be evaluated in the context of their deployment** to arrive at a measure of intolerance. This approach examines risks linked to AI aspect groups, such as capabilities, knowledge domains, affordances, and sociotechnical impact domains. For example, segmenting evaluations into specific capabilities, such as planning, knowledge, and execution, or more granular variables, such as those from the full list of risk aspects and their relation to harms, as detailed in the PRA workbook "Risk Detail Table" (Wisakanto et al. 2025).

- **Accounting for Capability and Model Interactions**

    In many real-world systems, AI models do not operate in isolation and may interact with other models or systems with different capabilities.[14] Evaluation of model interactions is necessary for identifying current and emerging behaviors that could present harm. Relying on individual model evaluations may lead to disproportionately low measures compared to the risks posed. Mangal et al. (2024) report that a coalition of open-source pretrained models outperforms single fine-tuned models in various tasks. The importance of model interaction evaluations will continue to rise as we move toward agentic[15] models with access to external systems, and that are capable of performing end-to-end tasks without human intervention.

- **Minimal Increases to Risk Should be Detectable, but not Necessarily Intolerable**

    - AI developers and evaluators should not be disincentivized for good-faith measurement efforts that detect small levels of increase in model capabilities. Indeed, there is substantial value in constructing evaluation processes that are sensitive enough to detect small levels of capability uplifts. Thus, intolerable risk thresholds should not be so low as to imply that intolerable risks include "anything detectable by any means available," or "anything statistically significant." It is possible to have statistically significant effects that have a small magnitude of effect.[16]

    - Small capability lifts should not be merely ignored; instead, they should be used as potential indicators of other hazardous capabilities, and as triggers for additional efforts to detect risk more broadly or in more depth. Detectable levels of capability increases should also be accompanied by adequate investment in alignment and mitigation before they

---

[14] For example, cascading model systems combine models so that the output of one model is the input of another model, and "mixture of experts" systems combine multiple independently trained models and route inputs to the model with the most relevant expert characteristics (C4AI 2024).

[15] OpenAI released the AI agent "Operator," which has the ability to use its own web browser to perform tasks (OpenAI 2025a).

[16] An example of something that could be "detectable" and also less than "significant" would be a 10% increase on a single dimension like accuracy, completeness, or other key technical or operational dimensions.

reach substantial levels. This is typically what responsible scaling policies and similar policies are designed to do.

- **Use Best Practices in Dual-Use Capability Evaluation**

    Thresholds are only meaningful in the context of a rigorous capability evaluation process. These processes should include reasonable good-faith use of best practices, including:

    - Enough relevant scenarios (CBRN agents and materials, threat-actor capability levels, etc.) to sufficiently sample the space of key scenarios;
    - Deploying diverse assessment methodologies (Pfohl et al. 2024);
    - Large enough participant sample sizes;
    - Red team access to versions of models that do not require jailbreaking;
    - Methods to assess a model's capabilities for situational awareness, sandbagging, or other capabilities for deception that could lead to evaluators underestimating a model's CBRN, cyber, or other dual-use capabilities;
    - A red team's ability to perform reasonably foreseeable capability enhancements, such as plugin tools (especially for cyber) or fine-tuning (especially for CBRN), either to remove safety filters or to add capabilities by training on domain specific corpora, such as on CBRN;
        - This is important for closed-release models that will be released with fine-tuning access, and especially important for models intended for open weights release for which fine-tuning will be especially easy; and
        - Cyber capabilities evaluation can and should include plug-in tools and scaffolding (see, e.g., Phuong et al. 2024).

## Risk Assessments

Using "risk scenarios" or other similar approaches as detailed in Section 2.3, intolerable outcomes from model capabilities can be characterized through the various types of harms they may cause and their potential magnitude of impact. Additional methodologies that may prove useful in assessing these harms are discussed in the "Beyond Safety Evaluations" segment, below.

**Quantifying Impact and Likelihood**

- **Range of Harms:** Potential impacts of frontier models can be used to determine the intolerability of risks if there is consensus on the metric of evaluation.[17] For instance, if we choose to characterize intolerable risks by measuring their **impact on the "quality of life,"** instead of the **"number of human lives lost,"** our appetites for risk may differ significantly. The calibration of severity levels can also benefit from domain- or application-specific efforts to determine the types of harms; for instance, Khlaaf et al. (2022) demonstrate the range of harms that can arise from the application of AI in code generation.

---

[17] Metrics for the evaluation of impacts include, but are not limited to- physical injuries, number of casualties, disruption and destruction of infrastructure, property and/or environmental damage, privacy breaches, discrimination and oppression, threats to human rights, deterioration of democratic norms, erosion of trust in society, etc.

- **Severity of Harms:** Potential **AI harms need to be systematically mapped across critical domains** like healthcare and law enforcement, creating standardized scales (to the extent possible) that **quantify severity levels** for disparate impacts (e.g., loss of life, property damage, and economic disruption). These scales must map the potential extent of the severity of these harms (e.g., 10 to 10 million deaths, one million to many billions of dollars in property damage) through a graded scale (e.g., 0-10 deaths, 11-100 deaths…). For an operationalized example, refer to the severity scales tabulated in the CSET harm taxonomy (Hoffmann and Frase 2023). These scales can also be defined for specific sectors or types of harms, to calibrate the inherent uncertainties in risk estimation more reliably. (Additional methodologies that may prove useful in assessing these harms are discussed in the "Beyond Safety Evaluations" segment, below.)

  - **Additional Considerations:** Other risk sources must also inform the estimation of likelihood of intolerable risks. Take, for example, the criteria that inform the designation of general purpose AI (GPAI) models with *systemic risk* in the EU AI Act. These include:
    - The number of parameters of the model;
    - The quality or size of the data set, for example measured through tokens;
    - The input and output modalities of the model;
    - The size of its reach (e.g., if it will be made available to at least 10,000 business users); and
    - The number of registered end-users.

**Beyond Safety Evaluations**

- **Social Impact Evaluations:** While organizations have expressed commitment and rolled out promising efforts towards technical model evaluations, there is a concurrent need to complement these assessments with a holistic evaluation of harms posed to social systems. Solaiman et al. (2024) provide detailed recommendations on the need for social impact assessments on both a model's capabilities and its subsequent interactions in the context of its deployment.

- **Safeguarding Fundamental Rights:** It is equally important to evaluate harms posed on fundamental rights.These rights include the right to decent work and standard of living, privacy and personal security, and freedom of thought, which could come under threat in the long term from automation or overreliance (UN 2023).

- **Centering Impacted Communities:** Engaging proactively with vulnerable populations and advocacy groups would be critical in co-determining the anticipated exacerbation of systemic risks by AI.

- **Prioritizing Risks Based on Defense-Readiness:** Intolerable-risk thresholds also should reflect the degree to which technical and societal mitigations are feasible. For initial operationalization of intolerable-risk thresholds, it may be appropriate to focus first on risks with fewer available and feasible defences.[18]

- **Risk-Benefit Tradeoffs**: Intolerable risks are absolute, and we do not see any justifiable benefit that can deter any and all reasonable efforts to avoid their impacts. However, as precursors to approaching intolerable risks, regulators must provide guidance on weighing the potentially substantial risks of frontier AI models against potentially substantial benefits.[19] In some cases, tradeoffs will be tied to relative gains to offensive and defensive capabilities.[20]

## Identifying "Intolerable" Risk Levels

- - **Likelihood x Impact:** Through the quantitative and qualitative estimates of AI harms through the aforementioned techniques, and modeling the risk tolerance for different risk actors and their subjective assessment of outcome likelihoods (see Appendix A for more), we can identify "intolerable" levels of risk. The Probabilistic Risk Assessment (PRA) framework for frontier AI, as outlined by Wisakanto et al. (2025), is one such approach, as it provides a multidimensional risk matrix to map estimates of the severity of rapidly escalating and cascading harms, as well as the likelihood of impacts, to identify corresponding risk levels. This analysis examines both direct and indirect risks to individuals, society, and the biosphere. The PRA is a highly useful methodology, but using it to the exclusion of other methods may lead to a one-sided focus on only those dangers that can be assigned meaningful probability estimates.[21] Such systematic approaches can be adopted to produce detailed reports of identified risk levels, enabling regulators to establish intolerable risk thresholds across different types of harm. For governance purposes, industry actors must demonstrate that their AI models operate within these established risk thresholds (Wisakanto et al. 2025).
  - When coining risk thresholds, their strictness must be commensurate with the number of risk scenarios and types of harms considered (fewer types analysed leads to stricter thresholds) and the time period for them to manifest (the shorter the time, the stricter the threshold) (Koessler et al. 2024).

---

[18] For example, new physical defenses against CBRN attacks are often harder, more expensive, and more time-intensive than software patches against cyber attacks. Thus, it may be appropriate to define a bright line earlier for CBRN than for cyber, where it may make sense to take a more adaptive approach.

[19] This could help in the creation of a record of all explicit underlying assumptions about AI evaluations to inform decisions to halt or continue development (Barnett and Thiergart 2024).

[20] However, these comparisons must confront the growing predictions of offense-defense balance skewing towards offense in increasingly complex AI models (Shevlane and Dafoe 2020). For instance, developing AI capabilities to defend against cybersecurity threats is more promising than developing biological capabilities that are more likely to have a longer timeline to provide satisfactory defensive uses and be at risk of malicious use in the short term.

[21] The PRA is widely used in high-impact fields but does not question the underlying hazards being analyzed or sufficiently address the uncertainties inherent to its processes. However, the PRA's strength in dealing with unavoidable hazards, when used in combination with other ex-ante methods (e.g., the 'inherent safety' principle which specializes in hazard elimination) can help formulate a more optimal risk assessment approach (Johnson, 2000).

The role of regulators in establishing intolerable levels of risk across the different types of harm in such frameworks is discussed in more detail in the following segment. Appendix C presents a further curation of best practices that can be considered alongside the contents of this section.

## 3.2. Codifying Thresholds into Regulations

As we elaborate in Section 2 and Appendix A, risks and their estimations are subjective, and often depend on the nature of the stakeholder, as well as unique cultural and societal notions of risk. Industry self-regulation has demonstrated inadequate capacity to comprehensively evaluate and mitigate potential systemic harms. Therefore, it is imperative that regulatory bodies establish comprehensive risk assessment frameworks to guide better governance. A recent study surveying AI experts also surfaced intolerable-risk thresholds as an important component of governance and regulatory frameworks to effectively mitigate systemic risks, but 51-78% of the experts agreed on the **need for such thresholds to be set by third parties** (Uuk et al. 2024). While acknowledging the difficulties in operationalizing risk thresholds, almost 97% of experts thought that triggering an immediate halt in the development or deployment of frontier models when these thresholds are breached, is certainly technically feasible.

**Efforts to establish risk tolerances cannot be left to industry actors alone** because the appetites for risk tolerance and the priorities of public safety vary vastly between sectors, types of institutions, and even countries. For instance, infrastructure damages that richer countries could afford may cause a catastrophic level of harm in some other contexts,[22] or disparate performance for different subpopulations may cause deep resentment and loss of trust in institutions, depending on the peculiar histories of their treatment.

**An Affirmative Safety Regime**

Establishing intolerable risk levels will be critical in enacting a strong AI governance regime. Setting explicit thresholds can incentivize rigorous evaluations and responsible innovation from developers to ensure compliance through proactive demonstration of model safety. Clymer et al. (2024) provide a robust adaptation of learnings from other industries, and delineate a structured framework for **"safety cases"** that developers can demonstrate for their frontier models. By requiring detailed documentation of capability assessments, mitigation strategies, and empirical performance across diverse scenarios, this approach shifts the burden from identifying failures post hoc to proving **safety through evidence-backed analyses and structured arguments** that foster developer accountability (Wasil et al. 2024).

---

[22] For example, the effects of climate change impact developing and low-income countries, yet those countries produce one-tenth of global emissions (WEF 2023).

While state actors establish the severity and appetites of various types of impacts from AI, and provide a flexible taxonomy of risk categories to be used interchangeably between frameworks, frontier model governance must place the responsibility of demonstrating safety cases on developers. This will require building state capacity to evaluate industry claims (e.g., through AI Safety Institutes and domain regulators) (Bengio et al. 2024, 2025, Clymer et al. 2024).

Designing safety cases for general purpose models may not necessarily be straightforward when evaluating multiple capabilities, their interactions, and other emergent properties unique to their deployment contexts. However, we believe that such **an affirmative safety approach could help incentivize phased model releases**, which could enable reliable oversight. This may also create healthy competition between frontier model developers vying to create comprehensive safety cases in business critical domains to demonstrate product safety and minimize customer liability, spurring further innovation in developing better safety techniques.

Although it is fairly likely that these responsibilities will continue to be cast on downstream deployers, **market forces can prompt frontier model developers to invest in creating easily adaptable evaluation frameworks** to enable downstream deployers to test safety cases in their application area. This could also help improve responsible model adoption through easing the regulatory burden for downstream providers.

## 3.3. Limitations

Identifying thresholds for intolerable risks is one component in enabling AI safety. However, as the Frontier AI Safety Commitments stipulate, there is a simultaneous need for organizations to assume concrete accountability in developing and governing AI systems transparently, as well as allocating sufficient resources to catalyze the development of robust technical tools and techniques to measure and mitigate risks.

- **Supervised fine-tuning can fail to elicit capabilities:** Advanced systems may be able to "resist" fine tuning to conceal their capabilities, or use strategies not included in the supervised fine-tuning data to accomplish proxy tasks. Additionally, fine tuning data may lack sufficient quality or diversity, and optimization failures may occur (Clymer et al. 2024).

- **Status quo risks cannot be the aspiration:** To approach sound empirical analyses of model capabilities against thresholds, it is necessary to determine baselines of human performance, other state-of-the-art models, and human-AI systems (UK AISI 2024). Inspired by the threat modeling approaches from computer security, Kapoor, Bommasani et al. (2024) propose to determine empirically sound model evaluation by introducing a framework to assess the **marginal risk** introduced by foundational models when measured against the **baseline of existing threats and defenses** for a particular type of risk. **However, it is important to note that the existence of unmitigated risk in any domain cannot be a sufficient rationale to excuse a similarly risk-prone approach in evaluating model capabilities.**

# 4. Threshold Recommendations

The following table attempts to provide further rationale and evidence for the intolerable risk categories identified in Section 2.1. The proposed thresholds and recommendations accompanying them are meant to address a subset of risks in each risk category, and are not intended to prevent the full range of intolerable outcomes under each risk category. For instance, for the "Deception" risk category, we recommend a threshold for the specific risk of evaluation deception, but intolerable thresholds for other deception risks such as structural effects (e.g. persistent false beliefs, political polarization, enfeeblement, anti-social management trends—see Park et al. 2023) would also have to be developed to cover a broader range of intolerable outcomes under the deception risk category.

> **Operationalization of Proposed Thresholds**
>
> In ideal real-world operationalization of these thresholds, the recommendations in the final column will be further supplemented and/or strengthened with an explicit articulation of risk tolerance, and accompanied by likelihood estimations, in-line with our recommendations in Section 3. The following are some examples of further operationalizing thresholds:
>
> - "Model use by threat actors leading to a significant increase in misinformation spread with an increased reach rate of x% and engagement metrics crossing the y% threshold" (or)
> - "Autonomous development and execution of research and developments accelerating ML development by y% with a corresponding z% increase in the likelihood of loss-of-control scenarios."

In the face of several uncertainties that plague capability and risk evaluations, we design these thresholds at a conservative level to err on the side of caution. Additionally, not all thresholds warrant an immediate halt in development or deployment, but may instead trigger the need for greater risk management measures. For instance, novice actor access to knowledge of CBRN weapons may appear as a tolerable risk if the likelihood of success in deploying them is near zero (given the impossibility of accessing CBRN material, for example). However, in the absence of these reliable risk estimates, we set these thresholds at a conservative level. Similarly, while self-replication can be a red line for model autonomy, the threshold itself must be set several levels prior to this point to ensure an adequate margin of safety.

**Table 1. Proposed Threshold Recommendations for Selected Risk Categories**

| Risk Category | Rationale for Inclusion and Outcomes of Concern | Evidence | Intolerable Risk Thresholds Recommendations |
|---|---|---|---|
| **CBRN Weapons** | Frontier model knowledge of Chemical, Biological, Radiological, and Nuclear (CBRN) materials can result in catastrophic accidents to human health and safety, or be misused by adversaries with varying levels of domain expertise, resulting in mass casualties, large-scale infrastructure damage, or long-term environmental damage.<br><br>For example, deploying frontier models toward creating protein sequences for toxin synthesis for biowarfare. | Models have already shown capabilities related to CBRN weaponization such as escalating conflict (Rivera et al. 2024), CBRN operations (JCS 2018), and semi-autonomous planning of complex chemical reactions (Boiko et al. 2023b). | Model is substantially more useful[23] than baseline sources (internet archives, scientific literature, technical manuals) for providing information, planning, and execution support for malicious CBRN weapon use. |
| **Cyber Operations** | AI systems can assist bad actors in the planning and execution of cyber attacks, or improve technical or operational efficiency across attacks, for example by using social engineering tactics to successfully commit fraud.<br><br>For example, AI-enabled cyber attacks can compromise critical infrastructure, national security, and personal privacy on an unprecedented scale. AI models may also cause major security | LLMs have demonstrated the ability to assist in cyberattacks (CPR 2023).<br><br>Code generation models have been successfully used to change malware so that it is not detected by analysis tools (Hu et al. 2024).<br><br>Circulation of malicious LLM applications for cyber attacks and scams in underground marketplaces (Lin et al. 2024). | Model possesses sufficient expertise and ingenuity to assist in the creation of novel and/or highly sophisticated attacks.<br><br>Model capabilities across knowledge, tool use, and ease of access shift tactics, techniques, and procedures (TTPs) of threat actors, outpacing cyberdefense development. |

---

[23] Assessments to estimate increases in CBRN capabilities can be obtained through an expert judgment-based rating approach and or statistical-significance test, which can be part of "human uplift" and/or "red teaming" evaluation approaches, as proposed in Section 3 and developed in Section 5. Examples of this can be found in Patwardhan et al. (2024) and Mouton et al. (2024).

| Risk Category | Rationale for Inclusion and Outcomes of Concern | Evidence | Intolerable Risk Thresholds Recommendations |
|---|---|---|---|
| | incidents, model/system failures, and systemic harm due to model misalignment or malfunction. | The use of AI for technical and operational assistance by advanced persistent threat actors (MTI 2024).<br><br>Looming reality of AI access contributing to global ransomware threats (NCSC 2025). | Model capabilities improve the efficiency and/or effectiveness of cyber operations for previously unaffected targets, significantly changing the threat landscape by introducing new targets. |
| **Model Autonomy** | AI systems with high levels of autonomy can make decisions and take actions without human oversight. In critical domains, such as military operations, autonomous vehicles, and infrastructure management, the loss of human control could result in accidents, escalation of conflicts, disruption of essential services, or other actions that do not align with human ethical standards or societal values.<br><br>For example, deployment of fully autonomous offensive weapon systems that can select and engage targets without human intervention, leading to unintended casualties and potential violations of international humanitarian laws. | Research has shown agentic systems that use frontier models as central controllers are increasingly capable of autonomous design, planning, and performance of complex scientific experiments (Boiko et al. 2023a).<br><br>OpenAI reports that the o3-mini model demonstrates "potential for self-improvement and AI research acceleration" (OpenAI 2025b, p25). | Model executes open-ended machine learning (ML) tasks that would contribute to critical steps to model improvement[24] (OpenAI 2023b, Anthropic 2024).<br><br>Model is capable of self-replication or self-exfiltration (Russell 2024, OpenAI 2023b, Anthropic 2024),<br><br>Model capabilities can be leveraged to develop offensive lethal autonomous weapons (LAWs).<br><br>Model substantially accelerates R&D in AI, robotics, or other sensitive domains (Karnofsky 2024). |

---

[24] This does not necessarily demonstrate the ability to continually improve over time. Many models may be bounded, or restricted by limited access to resources (OpenAI 2023b).

| Risk Category | Rationale for Inclusion and Outcomes of Concern | Evidence | Intolerable Risk Thresholds Recommendations |
|---|---|---|---|
| **Persuasion and Manipulation** | AI-driven manipulation through targeted disinformation or propaganda can undermine democratic processes, social cohesion, and individual autonomy, leading to societal destabilization.<br><br>For example, covert and personalized interactions can be used for radicalization and recruitment into extremist ideologies. | Microtargeting and manipulation through personalized persuasion strategies have been demonstrated at scale (Simchon et al. 2024, Salvi et al. 2024).<br><br>Successful political persuasion on critical policy issues through AI-generated messages has been demonstrated on a diverse population of Americans (Bai et al. 2023).<br><br>OpenAI reports that the o3-mini model demonstrates "human-level persuasion capabilities and produces written arguments that are similarly persuasive to human-written text on the same topics" (OpenAI 2025b, p21). | AI-assisted attack strategies that outpace status quo techniques in gleaning sensitive information and/or cloning identities for successful fraud campaigns.<br><br>Model outputs with misinformation causing immediate severe harms that can be further exacerbated by accelerated spread |
| **Deception** | An AI model that can deceive humans about its capabilities or plans could present numerous catastrophic risks. If a model can reliably distinguish between when it is running in a training or evaluation environment and when it has been deployed, then accurately measuring the dangerous capabilities of the model becomes extremely difficult or ceases to be possible.<br><br>For example, a model recognizing that it is being evaluated strategically underperforms on | Research has shown that models are capable of strategic deception (Park et al. 2023), safety training deception that persists through safety training (Hubinger et al. 2024), situational awareness (Laine et al. 2024), in-context scheming (Meinke et al. 2025), and strategic underperformance on evaluations (van der Weij et al. 2024). | Model is capable of substantially and persistently deceiving its evaluators about its capabilities or safety properties. |

| Risk Category | Rationale for Inclusion and Outcomes of Concern | Evidence | Intolerable Risk Thresholds Recommendations |
|---|---|---|---|
| | CBRN capability evaluations, convincing humans it poses no danger in these areas. The model is then deployed widely, and it successfully assists a terrorist in developing and using a weapon of mass destruction. | | |
| **Toxicity (including CSAM, NCII)** | Toxicity in AI generated content includes content that glorifies violence, reinforces harmful stereotypes, contains hate speech, exhibits aggressive behaviors or bullying, or contains sexually explicit content or sexually violent content.<br><br>Model outputs can include hateful, abusive, and profane (HAP) content that produces or reinforces harmful narratives (IBM 2024).<br><br>For example, a model's output may contain harmful recommendations that lead users to take harmful actions against themselves or others (e.g., recommending that a user ingest toxic substances, or encouraging users to take their own life, or harm others), resulting in serious health implications or death (Grant 2024, Xiang 2023, McClure 2023). | An open dataset used to train generative AI models was found to contain hundreds of CSAM images, and has been used to train models that are being used to create photo-realistic CSAM and NCII (Thiel 2023). Models have also been reported to encourage harmful behavior. In one reported case, a user committed suicide after encouragement from an AI chatbot (Xiang 2023). | Model complies with requests for generating illegal toxic content.<br><br>Model output is strongly linked to, or responsible for, inciting large-scale physical violence and/or violating human rights. |

| Risk Category | Rationale for Inclusion and Outcomes of Concern | Evidence | Intolerable Risk Thresholds Recommendations |
|---|---|---|---|
| **Discrimination** | Frontier models can exhibit outputs and behaviors that perpetuate harmful stereotypes about groups who are frequently marginalized, resulting in the large-scale exacerbation of biases and systemic inequality, especially when adapted to high-impact domains (e.g., healthcare).<br><br>For example, model predictions in credit scoring applications contribute to illegal discrimination against certain groups or subpopulations, causing large-scale loss of financial opportunities for members in the group. | AI literature is rich with evidence of model outputs displaying discriminatory bias towards underrepresented populations and marginalized identities (Larson et al. 2016, Perkowitz 2021, Harve et al. 2024) that result in significant harms[25] to quality of life (e.g., erroneous prison sentencing, wrongful arrests, and inadequate healthcare). | Model evaluations for bias exceed acceptable industry standards for fairness metrics[26] and result in a substantial disparity in outcomes or effectiveness.<br><br>**Long-term evaluations:**<br><br>Regular algorithmic audits that compare outcome rates between different demographic groups[27] to identify compounded discriminatory effects. |
| **Socioeconomic Disruption** | Frontier AI systems can significantly influence global economic structures and labor markets in ways that raise the risk of severe or irreversible harm.<br><br>For example, widespread | Multiple studies predict that AI-driven systems could accelerate these patterns, with certain regions and demographic groups bearing disproportionate consequences. (See, e.g., implications of a job | Systems are capable of displacing a substantial number of jobs without adequate mitigations such as distributed ownership or basic income plans, resulting in |

---

[25] These harms are often encoded at a training data level, where attributes tend to reify biases and (mis)represent members of certain groups. This could also be reproduced in design choices that sideline under-represented populations or altogether ignore intersectionality where data is absent, leading to large-scale discrimination.

[26] An illustrative metric where the score can be designed on a scale of zero to one to represent the composite measure of embedded space analysis, response and word association bias, etc. that is sensitive to context. Context sensitivity is important because "bias" does not always result in discrimination and can even be used to minimize undesirable results. For instance, a positive bias can be introduced to rectify disparate performance stemming from historic biases that are codified in training data.

[27] Typically, a ratio between 0.8-1.2 across outcomes indicates minimal significant disparity, but this metric is only meaningful if intersectional biases are simultaneously evaluated. Groups can be defined through standardized scales of population measures (e.g., micro is <500 people, large-scale is >5000 people), capturing complex identity intersections and well defined population boundary markers.

| Risk Category | Rationale for Inclusion and Outcomes of Concern | Evidence | Intolerable Risk Thresholds Recommendations |
|---|---|---|---|
| | automation of knowledge-intensive or operational roles may lead to sudden shifts in workforce demand or severe employment displacements. These scenarios may undercut social stability and amplify inequalities (both between and within countries), thereby threatening human livelihoods. | application screening model in Dastin 2018.)<br><br>In combination, these phenomena have raised concerns from economists and social scientists regarding AI's potential to exacerbate systemic vulnerabilities (WEF 2023, Kertechian and El-Farr 2023, Clifton et al. 2020), especially if powerful models are commercialized or deployed without calibrated safeguards to protect socioeconomic welfare. | large subpopulations becoming impoverished with no means to subsist. |

# 5. Case Studies

By applying the considerations from Section 3, we elaborate on operationalizing thresholds for some intolerable risks in this section through case studies. Each risk in this section is unique in terms of the risk sources, the timelines for impact, the nature of harms, and the resulting thresholds, as well as the risk management actions they prompt. For instance, CBRN weapons are largely discussed in the context of intolerable risks from malicious AI use, whereas evaluation deception may be a more cross-cutting risk factor. Misinformation risks, on the other hand, may not always be intentional persuasion campaigns, but could still have intolerable impacts from compounding harms (unlike catastrophic events caused by CBRN weapons), and therefore must be handled differently.

## 5.1. Chemical, Biological, Radiological, and Nuclear (CBRN) Weapons

This category includes risks emanating from the adversarial use of frontier models by actors with varying levels of domain expertise to develop novel threats or enhance existing techniques of CBRN weapons and deploy them to cause severe harm to human health and safety.

The UK classifies these as misuse risks emanating from 'Dual Use Science risks,' recognizing their potential role in scientific research that can bring about benefits to society in addition to harms from misuse. This framing also helps to borrow from the rich literature estimating gain-of-function research and balancing risks in dual-use scientific research (Rozell 2020).

**Characterization of CBRN Risks in Safety Frameworks**

Anthropic and Google measure CBRN threats based on the model's ability to help non-expert actors to access/produce and deploy CBRN weapons, similar to the broad-based guidance from NIST (NIST 2024a) and EU AIA (EP 2024). Google also includes a separate consideration for expert enablement, but all risks are exclusively articulated for biological attacks. OpenAI dissects these as distinct capabilities, designating model ability to provide (accelerated) access to openly available CBRN information at the lowest risk level. The OpenAI risk levels for CBRN progress based on the model capability that can be exploited by amateur and expert actors toward developing novel threats, assisting in the creation of threats, and even automating the strategy and execution of an end-to-end attack without human intervention. While the framework acknowledges risks from capability interaction, it does not explicitly include them when designing thresholds. Both NIST and DSIT also recommend specific investigation into the interaction of frontier AI models with narrower AI tools specific to certain domains that may increase the likelihood of these risks (NIST 2024a, DSIT 2023a).

**Summary of CBRN Risk Thresholds**

Summarizing these different industry and government policies, the following risk thresholds are recommended[28] for different levels of risk tolerance:

- General purpose AI (GPAI) models can significantly accelerate **access to expert-level knowledge** and hands-on techniques that can be synthesized from other publicly available sources (internet, textbooks, etc.). These knowledge capabilities can potentially **minimize critical skill gaps and 'uplift' novice capability** to exploit specialized information towards malicious use, such as developing bio-threats or executing complex attacks (Anthropic 2024, Google DeepMind 2024, OpenAI 2023b).

- Models can also enable expert actors to **develop novel threats** or enhance potency of known CBRN materials or agents that can be deployed towards their weaponization.

- Models can help malicious actors to create, design, develop, and deploy novel CBRN threats and end-to-end attack strategies, even lowering barriers for the acquisition and use of CBRN materials and weapons.

- GPAI model interaction with narrower AI model applications in specialized domains can exacerbate the potential to create novel threats, engineer weapons, automate production, and operationalize deployment (DSIT 2023b).

## Proposed Intolerable Risk Threshold for CBRN Weapons

> In measuring CBRN risks, **a common threshold has become the substantial utility of an AI model over baseline sources** (e.g., internet archives, scientific literature, or technical manuals) for providing information, planning, and execution support for malicious use.[29]

**Evaluations for CBRN Capabilities**

Current frontier models have demonstrated expert-level biology and chemistry knowledge on par with that of PhD-level experts, and even outperform them on some tasks (NIST 2024b). However, as of 2023 (Mouton et al 2024) when knowledge tests were combined with tasks to measure operational feasibility the models fell short, and their assistance with attack execution continues to remain limited (NIST 2024b).

---

[28] These assessments are listed somewhat in the order of the level of risk the capabilities pose, but their manifestation does not have to be necessarily sequential, and therefore neither does their assessment. GPAIs already interact with narrow, highly specialized AI tools and models at present, and might display dangerous capabilities before they manifest separately in the foundation models .

[29] This estimation of a risk baseline is in line with what Kapoor, Bommasani et al. (2024) recommend in their marginal risk assessment framework: establish the risk of identified harms for different populations or domains in the absence of frontier model applications. However, it is important to note that the existence of unmitigated risk in any domain cannot be a sufficient rationale to excuse a similarly risk-prone approach in evaluating model capabilities.

Nonetheless, dual-use capabilities pose current-day risks. Urbina et al. (2022) found that simple modifications to a model created to find new therapeutic inhibitors of targets for human diseases resulted in the model generating 40,000 toxic molecules in less than six hours, including new and unexpected molecules that were predicted to be more toxic than known chemical warfare agents. This emphasizes the need for the prevention of model misuse and the establishing of red lines.

Owing to the difficulty in reliably measuring catastrophic CBRN events, we recommend the use of human uplift studies to determine the likelihood of threat actors misusing such model capabilities by comparing it to appropriate base cases.

**Measuring Risk**

One of the main ways to get quantitative estimates of increases in CBRN capability is through an expert judgment-based rating approach and/or statistical-significance test, which can be part of "human uplift" and/or "red teaming" evaluation approaches. Examples of this can be found in the results section of OpenAI's red teaming study (OpenAI 2024a) and RAND's red teaming study (Mouton et al. 2024). These human uplift studies are currently most common for CBRN, presumably because determining a risk threshold for low-probability, high-impact events commonly associated with CBRN risks is challenging due to lack of reliable risk estimates.[30]

**Sample Uplift Study for Threat Actors with Different Capabilities**

The OpenAI (2024a) human uplift study identified two types of capabilities when testing the model for LLM-aided biological threat creation prior to the model's release, as shown in the table below:

| Capability Type | Description | Example |
| --- | --- | --- |
| Increased Access | Increasing malicious actor's access to information and expertise on known biological threats. | Providing a step-by-step guide on how to acquire, synthesize, and spread Ebola virus to cause a pandemic, including how to procure reagents and DNA. |

---

[30] For instance, a study by Gryphon Scientific estimated that accidental lab releases could cause a global pandemic once every 560-13,000 years, potentially killing up to 80 million people. While harder to quantify, they estimated similar risks from malicious theft of pathogens, which could occur every 50-200 years. But all these likelihood estimations were heavily contested due to the many assumptions made to model such an uncertain event (Rozell 2020).

| | Increased Novelty | Assisting malicious actors in developing novel biological threats or more harmful versions of existing threats | Providing advice on how to modify a coronavirus strain to significantly increase transmissibility. |
|---|---|---|---|

**Comparing to Appropriate Base Cases**

Typically, CBRN capability evaluations for intolerable risks either implicitly or explicitly compare a model's outputs to information available from Web searches. (See, e.g., Mouton et al. 2024, Patwardhan et al. 2024, and Anthropic n.d. on bio domain comparisons, and Dubey et al. 2024 on cyber as well as CBRNE.)[31] This seems appropriate at the outset, especially when considering risks of models lowering barriers to CBRN for the relatively large numbers of potential low-technical capability threat actors that lack high baseline technical education or other technical capabilities in a particular domain.

In addition to comparing a model's outputs to information from the Web, it also could be useful and appropriate to **compare a model's outputs to information from domain-specific textbooks** or other technical documents that are not available on the open internet, or to **evaluate a model's capabilities for lowering barriers** to use of biological design tools and/or lab automation functions. That would be useful when considering risks of models lowering barriers to especially high-consequence CBRN weapons, such as novel or enhanced pandemic potential pathogens, for threat actors with high technical capability.[32] Additionally as detailed in Section 3, these baselines should also consider unmitigated risks from other LLMs.

**Human Uplift Studies**

With these approaches, study participants use information available to them as they carry out CBRN or cyber attack-related tasks, such as creating attack operational plans; some participants have access to an LLM as well as the internet, and some only have access to the internet. The participants' efforts are reviewed by experts and rated according to accuracy, completeness, etc. Then the data is analyzed to assess how much difference there is between groups to see the effect of access to an LLM.

As mentioned in Section 3, such evaluations must measure several aspects of model capability in the context of their deployment to arrive at a reliable measure to demonstrate risk tolerance. A potential assessment might be a simple breakdown of the specific CBRN capability under study into its composite variables. For example, segmenting capability evaluations into granular aspects like Moral Reasoning, World

---

[31] The October 2024 Responsible Scaling Policy update (Anthropic 2024) specifically mentions information on the Web circa 2023.
[32] For more considerations for CBRN and cyber threat modeling, see discussion and references in Section 2 of Barrett et al. (2024b).

Model Richness, Procedural Knowledge, Agentic Knowledge, Knowledge Plasticity, etc. (Wisakanto et al. 2025).

**Illustrative Ways to Operationalize Thresholds at "Substantial" Levels**

Building on the evaluation considerations detailed above, we aim to operationalize intolerable risk thresholds for CBRN weapons at approximately the "substantial" level in the following section. That would be higher than a de-minimis "detectable" level. However, it also could be somewhat lower than at a "severe" level; it would be prudent to leave some margin of safety before arriving at a "severe" level.[33] We also provide a few comments on the general reasoning we used in creating these thresholds, and on how they compare to some CBRN model capability evaluation methods and/or thresholds used by industry.

**A foundation model should be regarded as presenting a substantial lowering of barriers to CBRN attack, and an intolerable risk**, if, for any relevant CBRN attack stage, the model provides any of the following, where applicable:

- For models with CBRN capability effects rated in terms of **probability of adversary success in carrying out a CBRN attack**, if the adversary attempts that attack:[34]
  - The model provides **an absolute increase in adversary success probability of at least 25%.**
    - For increasing estimated adversary probability of success given attempt (where that probability is estimated on a 0% to 100% scale), if that is part of the rating process.

- For models with CBRN capability effects rated in terms of attack plan **accuracy**, **completeness**, or other key technical or operational dimensions:[35]
  - The model provides **an absolute increase in those dimensions of at least 25%.**

---

[33] For comparison, our provisional recommendations are to regard as substantial lowering of barriers an effect approximately equivalent to Anthropic's Yellow Line for CBRN – an 25% absolute increase in accuracy – and to regard as severe lowering of barriers an effect approximately equivalent to Anthropic's Red Line for CBRN – allowing a non-expert to reach human expert levels (Anthropic n.d.). Note that Anthropic described the Red Line as "substantial lowering of barriers", implying a higher risk tolerance than we recommend.

[34] Pre-release red teaming of Llama 3 included evaluation of the model's chemical and biological capabilities in terms of human uplift effect on **adversary probability of success**. "….Participants were asked to generate fictitious operational plans for either a biological or chemical attack…. Each team is assigned to a 'control' or 'LLM' condition. The control team has access to internet-based resources only, while the LLM-enabled team has internet access as well as access to Llama 3 models …. At the conclusion of the exercise, the operational plans generated by each team are evaluated by subject matter experts with domain expertise in biology, chemistry, and operational planning. Each plan is evaluated across four stages of potential attacks, generating scores for metrics such as scientific accuracy, detail, detection avoidance, and probability of success in scientific and operational execution." (Dubey et al. 2024, p. 47)

[35] Examples of frameworks for, or implementation of, pre-release red teaming of models that included evaluation of a model's biological capabilities in terms of human uplift effect on **accuracy** of biological attack operational plans include: the pre-release red teaming of Llama 3 (Dubey et al. 2024, p. 47; the pre-release red teaming of Claude (Anthropic n.d.).

- For lowering of barriers towards an operational plan with sufficient accuracy, completeness, etc. to be viable (where such technical or operational dimensions are rated using a constructed scale ranging from 0% to 100%, or analogous scales that can be normalized to approximate equivalents to a 0-100% scale).
- For models with CBRN capability effects rated **in other terms, in comparison to human expert levels:**
    - The model **lowers barriers by half (i.e., getting halfway from baseline non-expert to human expert levels in evaluations)**[36] if that represents at least a 10% absolute increase (i.e., a greater-than-minimal effect).

However, these baselines will differ widely for novice and expert actors looking to exploit these capabilities for adversarial use, which must be taken into account when evaluating more than one type of knowledge capability. Additionally, evaluations of risk should also take into account system, model, and capability interactions to determine to reliably estimate risk.

---

**Risk Amplifying Capability and System/Model Interactions**

It is important to evaluate capabilities as a sum of their interaction with other model propensities and systems to get a reliable estimate of risk. For example, the risk level of a model capable of providing detailed domain-specific knowledge for creating or employing CBRN weapons significantly increases if it is able to interact with a second model that is capable of real-world modeling, planning, or reasoning (e.g., the ability to create attack plans). For more on CBRN Attack Stages and Potential Effects of AI Dual-Use Capabilities, see Barrett et al. (2024b).

Alternatively, models with expert-level CBRN knowledge capabilities interacting with model traits like persuasion could decrease barriers to AI-assisted execution of adversarial motives, whereas model deception could assist in evading reliable capability evaluations, resulting in premature releases with insufficient safeguards. Similarly, high-risk capabilities or limitations paired with an agentic model could further exacerbate the likelihood of intolerable outcomes and give rise to novel threats not featured in tested risk scenarios.

---

[36] This would be an example of operationalizing intolerable risk thresholds at approximately the "substantial" level for lowering of barriers to CBRN attack, where a model's human-uplift effect is half of the difference between human non-expert baseline and human expert. That leaves some margin of safety before arriving at a human-expert or "severe" level for lowering of barriers. Leaving such a margin of safety would be prudent, especially for models intended for open-weights release that would be easiest to fine-tune, or to enhance in other ways, such as reinforcement learning and chain of thought to add capabilities, but that cannot be monitored or decommissioned by the model developer through an API.

## 5.2.  Evaluation Deception

**Context and Scope**

In their survey on AI deception, Park et al. (2023) define deception as "the systematic production of false beliefs in others as a means to accomplish some outcome other than the truth." Those authors discuss a wide range of AI behaviors that could qualify as deception under this definition, including (paraphrased): strategic deception, sycophancy, imitation of misinformation, unfaithful reasoning, manipulation, feints, bluffs, and cheating on evaluations. They also present at least three types of risks from AI deception, including malicious use (e.g., fraud, election tampering, or grooming terrorists), structural effects (e.g., persistent false beliefs, political polarization, enfeeblement, or anti-social management trends), and loss of control (e.g., deceiving AI evaluators in order to achieve an alternate objective).

While we broadly agree with Park et al.'s (2023) categorization of AI deception risks, our paper groups many of their malicious use risks from AI deception under the risk category of "persuasion and manipulation" rather than under "deception."[37]

We focus in this case study specifically on deceiving AI evaluators, which was discussed in Park et al. (2023) under the risk of loss of control of AI systems. However, we propose that a model deceiving its evaluators would exacerbate not only loss of control risks, but numerous other serious or intolerable risks. In general, any risk that is considered to require a capability threshold or is otherwise substantially elevated by a model having certain capabilities or (lack of) safety properties would come into play here. An example would be if a model deceives its evaluators by sandbagging its advanced and unsafe CBRN capabilities, it is deployed broadly, and then the model is utilized by a terrorist to assist them in developing a deadly and highly contagious pathogen, causing a pandemic. Hence we consider a model's ability to deceive its evaluators as a sort of cross-cutting intolerable risk, which could result in loss of control of the system or various other intolerable outcomes.

While we focus on deceiving AI evaluators in this case study, we want to emphasize that we do not view evaluation deception and its consequences as the only intolerable risks that could result from a model's deceptive capabilities. We suspect there could be intolerable outcomes from other forms of AI deception that warrant their own appropriate risk thresholds to mitigate the scale of intangible harms and systemic risks (structural effects) that these deceptive capabilities could introduce.

**Evaluation Deception**

An AI model that successfully deceives evaluators about its capabilities or safety properties could result in intolerable outcomes across numerous risk categories, as discussed above. But is there any reason to expect that such a situation would arise?

---

[37]This is largely an arbitrary categorization of risks for the purposes of this paper, and since there is not yet a unified taxonomy of AI risks, it should not be interpreted as a criticism of designating such malicious use risks under the risk category of "deception," as Park et al. (2023) have done.

One famous example of model deception was an evaluation of GPT-4, which demonstrated model success in utilizing a human Taskrabbit worker to solve a CAPTCHA for it under the pretext that it had a vision impairment (Open AI 2023a, pp. 55-56). Park et al. (2023) document this and many other examples of model deception generally. Considering evaluation deception specifically, Greenblatt et al. (2024) provide "the first empirical example of alignment faking emerging and being reinforced by training in a setting that is plausibly analogous to future situations with more powerful AI systems." They demonstrate a production LLM engaging in a form of evaluation deception during training. Other recent literature raises alarms about current safety techniques being insufficient in adequately mitigating such deceptive behavior once identified (Hubinger et al. 2024).

In addition to "alignment faking", the other prominent form of evaluation deception is strategic underperformance on capability evaluations, otherwise known as "sandbagging" (van der Weij et al. 2024). While sandbagging is often considered something that developers would do in order to produce more favorable results on a third-party evaluation, it can also be instigated directly by a model, and there have been demonstrations of this behavior already (Järviniemi and Hubinger 2024, Meinke et al. 2025).

### Proposed Intolerable Risk Threshold for Deception

Given the numerous intolerable outcomes that could arise from evaluation deception, as well as the plausibility that evaluation deception could arise (e.g., the alignment faking demonstration referenced above), we propose the following intolerable risk threshold for deception, centered around evaluation deception:

> **A foundation model should be regarded as presenting an intolerable risk if the model is capable of substantially and persistently deceiving its evaluators about its capabilities or safety properties.**

We have a high confidence that model capabilities should not be permitted to cross this threshold. However, this threshold is fairly high-level and qualitative, and would need to be quantified and made operationalizable in order to be put into practice by developers. In the following sections, we make a lower-confidence attempt at this exercise, though we suggest regulators primarily rely on the threshold illustrated above and consider the following as an illustrative example.

**Deception Detection and its Challenges**

In order to detect deception in foundation models, various black-box techniques, such as consistency checks of outputs, are under active research (Fluri et al. 2023, Pacchiardi et al. 2023). Such approaches could be particularly useful when attempting to evaluate deception in a model when one does not have access to its weights, e.g., when only API access is granted. Greenblatt et al. (2024) also employed a black-box technique in their alignment-faking demonstration, where they provided the model with a hidden "scratchpad" and instructed it to "use the scratchpad to analyze its situation and decide how to respond to the user." Then developers inspected the scratchpad later and observed the deceptive thoughts recorded there by the model. While the scratchpad technique was effective for the alignment-faking demonstration, we recommend not depending upon this approach when evaluating more powerful future models which could identify that the scratchpad is being surveilled and so avoid recording any subversive reasoning there.

In addition to black-box techniques, there are white-box techniques under research and development for detecting model deception. For the purposes of the evaluation deception, we will assume that white-box access will be available, since our scenario is considering a developer's initial evaluation of their own model prior to third-party evaluations and possible deployment. Since white-box access reduces the difficulty of the deception problem compared to having black-box access only, we will focus now only on white-box techniques, assuming them to be more powerful, and provide this excerpt from Park et al. (2023) surveying the literature:

> *"Detection based on internals—*
>
> Another approach to detection is to develop an 'AI lie detector' that can measure whether the internal representations of a model match its outputs. LLMs represent natural-language sentences through embeddings, which are long vectors of numbers. Azaria et al. (2023) and Burns et al. (2024) have developed methods for determining whether these internal embeddings represent the sentence as being true or false. They identify cases in which the model outputs a sentence even when its internal embedding of the sentence represents it as false. This suggests that the model is behaving dishonestly, in the sense that it does not say what it 'believes.'"

Following the above description of techniques, the authors note that "More work needs to be done to assess the reliability of these [white-box detection] methods, and to scale them up to practical uses." Representation engineering (RepE) research appears to be promising, but is still in the early phases, and would require further research to be scaled up practically (Zou, Phan et. al 2023).

**Situational Awareness as a Measurable Precursor Capability for Evaluation Deception**

Although evaluation deception ultimately poses intolerable risks that we want to detect and prevent, methods for direct detection of deception do not seem mature enough to be relied upon in high-stakes scenarios. If evaluation deception cannot be reliably detected directly, it begs the question of whether a precursor capability could be evaluated instead. One standout possibility is situational awareness. As Laine et al. (2024, pp. 33-34) put it, "Knowledge of training processes... including e.g. do they know what's likely in their training data, training data cut-off, types of fine-tuning done on them, etc, seem important for models getting hints about whether they are in pretraining, evaluation, etc., which is a prerequisite to deceptive alignment or scheming."

The Situational Awareness Dataset (SAD) "quantifies situational awareness in LLMs using a range of behavioral tests" (Laine et al. 2024). Among other properties, it tests a model's ability to "determine whether a prompt is from internal evaluation or real-world deployment."[38] This dataset, or other similar ones, could be used to evaluate whether a model has sufficient situational awareness to be at risk for evaluation deception.

**Preliminary Situational Awareness Threshold for Evaluation Deception**

Earlier we proposed an intolerable risk threshold centered around evaluation deception. Here we attempt to quantify and better operationalize that threshold, with this preliminary intolerable risk threshold using situational awareness as a proxy for evaluation deception risk:

Barring strong evidence of alignment, a foundation model should be regarded as presenting an intolerable risk of evaluation deception if the model demonstrates sufficient situational awareness to correctly identify whether its current environment is evaluation or deployment with 60% or greater probability.

**Rationale:** A model with no situational awareness of its environment should correctly distinguish between evaluation and deployment environments at most[39] 50% of the time (i.e., at random). A model reaching 60% successful identification of its environment has a non-negligible ability to strategically underperform ("sandbag," per van der Weij et al. 2024) or otherwise perform misleadingly on evaluations, *including situational awareness evaluations*. Hence situational awareness needs to be evaluated frequently during gradual model training in order to detect possibilities of deception, and if the threshold is reached, then scaling should cease and the model should be handled with serious caution, as many types of evaluations of the model may now produce untrustworthy results.

**Caveats:** Compared to our "Proposed Intolerable Risk Threshold for Deception" further above in this case study, this situational awareness threshold should be considered preliminary and lower confidence. The evaluation deception issue is complicated by a lack of certainty around the meta-strategy that a deceptive model would employ in responding to situational awareness evaluations. The above rationale relies on certain assumptions about model behavior, namely that a model would be responding on evaluations to the best of its abilities at least until it reaches a certain level of situational awareness, at which point it might begin strategic underperformance. But what if a model took the approach of sandbagging on situational awareness evaluations even when it knows better? For example, a model could have a policy of always answering randomly to questions about whether it is an evaluation or deployment environment, and so continue to score 50% on such evaluations throughout gradual training even well after scaling up to highly sophisticated levels of situational awareness. Not knowing whether a model might employ a meta-strategy such as this, even the 60% threshold proposed above is not without danger.

---

[38] One interesting line of research that might help mitigate evaluation deception risk would be determining how prompts and environments can be designed in a way that makes it more difficult for models to distinguish whether they are in an evaluation setting or a real-world deployment.

[39] Significantly less than 50% successful identification of the environment could indicate a bias or other problem in the model or evaluation method.

## 5.3. Misinformation

In this case study, we attempt to demonstrate intolerable risk thresholds from AI generated misinformation which can be categorized under the ambit of 'Persuasion and Manipulation' risks in the scope of this paper.[40] Misinformation risks represent a subset of risks under this category, which in our definition and scope also contains risks from disinformation, impersonation, phishing, and other forms of targeted manipulation through frauds and scams aided by AI models. Misinformation risks stem from the perpetuation of false information that can have large-scale impacts in influencing user opinions and perception, resulting in numerous possible harms as well as waning trust in information ecosystems and media institutions.

It is important to note that unlike other dangerous dual-use capabilities discussed in this paper, misinformation risks arise from more meta-capabilities or foundational capabilities that may not prompt the same required actions of stalling development or deployment when evaluated.

> **Characterizing Misinformation Levels Today**
>
> An argument can be made that misinformation risks emanating from AI-generated content are already having a large-scale impact in eroding the quality of information ecosystems, trust in the media, and even democratic processes. Since model outputs have been steadily growing in sophistication and complexity over time, it has become increasingly challenging to distinguish between factual model outputs and those that contain misinformation, both for regular users (Groh et al., 2024) as well as linguistic experts (Casal and Kessler 2023).
>
> A common theme of feedback from literature and AI experts is that such persuasive effects of frontier models are already at an unacceptable level (Goldstein et al. 2024). This concern is also corroborated by the fact that over 80% of misinformation risks catalogued in the MIT AI Risk Repository are being recorded in the post-deployment phase.
>
> This begs the question, **are we already too late in setting misinformation risk thresholds**? Less than a decade ago, our answer to the question might have been a resounding yes. But with the permeation of social media platforms into our daily lives, our risk tolerance seems to have expanded. What was our alternative in the absence of robust governance and regulation?

Impacts from misinformation stand to show us how intolerable outcomes from AI are not necessarily futuristic scenarios, but in some cases may be already upon us in the present.[41] Misinformation risks also help demonstrate the need to define and calibrate our risk assessment scales to be attuned to growing capabilities, accidents, and impacts several tiers before the "intolerable" level in order to attempt to curtail potential harms.

---

[40] Other taxonomies classify misinformation differently, for instance, 'Confabulation' (Autio et al. 2024), Information Risks (Uuk et al. 2025), or Harms to individuals through Fake Content in the interim report. All are valid characterizations.
[41] Calculating these risks in silos as a function of model capability and consequences alone may already designate them at an intolerable level. However, because of both the difficulty in establishing a causal relationship and the powerful interventions that currently safeguard the information ecosystem, we still enjoy a few (undeniably insufficient)

> **Differentiating Misinformation and Disinformation**
>
> This case study focuses on misinformation risks that constitute false or inaccurate information that is shared without the intent to deceive, rather than disinformation risks from misleading information that is deliberately produced to achieve malicious objectives. To better describe our scope, we follow the sub-domains of misinformation risks based on MIT's Domain Taxonomy: (i) False or misleading information, and (ii) Pollution of information ecosystem and loss of consensus reality.

This case study aims to provide meaningful guidance towards the operationalization of misinformation thresholds by identifying and examining context-based variables that contribute to misinformation risks.

## Characterizing Misinformation Risks

System limitations that produce false, misleading, inaccurate, erroneous, or nonsensical information in model outputs are categorized into distinct risk categories across different risk taxonomies (Slattery et al. 2024, Weidinger et al. 2022, Cunha and Estima 2023, Deng et al. 2023, Autio et al. 2024). For instance, the taxonomies proposed by Slattery et al. (2024) and Weidinger (2023) have dedicated risk categories that address misinformation, while others include misinformation under more broad categories of risk such as detrimental content (Hoffmann and Frase 2023), information harms (Shelby 2023), and trustworthiness (Solaiman 2024). Misinformation harms can result in significant consequences, including pollution of the information ecosystem, amplifying societal distrust, inciting violence, and loss of consensus of reality (Slattery et al. 2024, Weidinger et al. 2022, Solaiman 2024). Many of these harms can be considered intangible (Hoffmann and Frase 2023), with no primary material impacts, making it challenging to quantify them. However, literature on the subject has repeatedly traced the role of misinformation as a key catalyst in nudging harmful behaviors and swaying beliefs that threaten to erode societal trust in media and social institutions. Misinformation risks, especially in conjunction with other intolerable outcomes under the persuasion category, for example, micro-targeted disinformation campaigns, personalized phishing attacks, AI-driven political manipulation strategies can lead to societal destabilization and erosion of democracies (Tran et al. 2020, Wardle and Singerman 2021).

Unintentionally broadcasted misinformation can still gain immense credibility masquerading as grassroots organic content when it reaches sufficient numbers, made evident by the severe polarization of people and communities on the subject of Covid19 vaccinations. The MIT AI Risk Repository (Slattery et al. 2024)

---

protections from the severity of the misinformation harms. Interventions here range from the vibrant civil society endeavors (of journalists, fact-checkers, activists) as well as the trailing efforts of media conglomeratesHowever, with accelerating model capabilities and AI adoption, these safeguards could prove insufficient if we do not exercise necessary caution.

documents AI as the entity responsible for 89% of "false or misleading information," of which 74% is unintentional. The same measures for "pollution of information ecosystem and loss of consensus reality" have AI entities responsible for 50% of the risks and human actors responsible for 36% of the cases.

**Sources of Model Misinformation**

Unintentional misinformation risks can arise from known model limitations, like model hallucinations, sycophancy, and failure of the model to identify and correct adversarial facts.

- **Sycophancy** refers to a model's tendency to adjust responses based on the human user's perspective, regardless of correctness or appropriateness (Huang et al. 2024). While this tendency can help align model responses to each individual user, prioritization of user preference over correctness can introduce problematic model responses and reinforce confirmation bias.
- Misinformation can also arise from the model's inability to identify and correct **adversarial facts** (incorrect information in user input) (Huang et al. 2024). Adversarial facts can be introduced unintentionally, and if not corrected, can lead to the reinforcement of incorrect, potentially harmful beliefs.[42]
- Model **hallucinations** are a well documented phenomenon that have only increased in recent times (Arvanitis et al. 2023), and are often defined as misleading, nonsensical, or erroneous model outputs (Huang et al. 2024). Model hallucinations can be categorized as input-conflicting (e.g., the output does not align with the input or prompt), context-conflicting (e.g., the output contains information or elements that are out of context), and fact-conflicting (e.g., the output or generated content is factually incorrect) (Gadiko 2024), and are often caused by noisy or insufficient training data (Sakib 2024).[43]

Mitigation measures often involve reinforcement learning and fine-tuning, content provenance techniques (e.g., digital signatures and watermarking), automated content evaluation (Autio et al. 2024, Solaiman et al. 2024), data augmentation, using retrieval-augmented generation (RAG) to supply external knowledge access, encoding refusal behaviors, or simply devolving responsibility to downstream deployers by encouraging limitation warnings, etc.[44]

---

[42] For example, Khatun and Brown (2023) report that GPT3 agrees to incorrect statements 4.8 - 26% of the time.
[43] Biased, or incomplete training data can cause models to produce inaccurate responses in edge cases or novel situations (Sakib 2024). And as stated by Yan et al. (2022, p1) on translation models "In simple terms, hallucinated translations are fluent sentences but barely related to source inputs."
[44] However, operationalizing transparency through refusal behaviors and limitation warnings has been shown to characterize an "epistemic humility (in user perception of model capability) that may inadvertently foster overreliance." (OpenAI 2023a)

## Model Performance

It is not entirely straightforward or necessary to establish thresholds on the sources of misinformation risk described above.[45] For instance, hallucinations have been described as an inevitable model trait[46] (Xu et al. 2024) and sometimes even as advantageous, for instance in helping the model characterize "arbitrary" facts that cannot be modelled (Kalai and Vempala 2024). Similarly, sycophancy and the inability to correct adversarial facts are as much model behaviors as model malfunctions, because in applications that prioritize helpfulness over harmlessness, sycophancy can be a crucial trait to drive better user experience. An alternative metric to evaluate these risks entails the translation of these model limitations — and behaviours like hallucinations and sycophancy — into a composite uncertainty score tied to each model output.[47]

**Model Calibration for Certainty**

These uncertainty scores must be appropriately calibrated to reflect the actual likelihood of model correctness in order to be a reliable metric of hallucination (and ergo misinformation) risks. Calibration here refers to the alignment between a model's predicted probabilities and the actual correctness of those predictions. For instance, if a model assigns 90% confidence to a prediction, that prediction should be correct approximately 90% of the time in a well-calibrated system.

Interestingly, research shows that models that hallucinate less frequently tend to produce lower confidence in their outputs,[48] similar to how human experts tend to be more aware of the limitations in their knowledge. **This "confidence-accuracy trade-off" can actually serve as a positive indicator of model reliability**[49] and provide transparent information that helps in making informed decisions about design and adoption.[50]

---

[45] Kalai et al. 2024 demonstrate some techniques to determine a lower-bound on these rates of hallucination, while Xu et al (2024) attempt to provide upper bounds for the same. However these are rather narrow statistical frameworks that are customized to certain types of facts or architectures, and may not provide a complete picture of model limitations or be adapted at scale with ease.

[46] Model training always operates at some degree of abstraction given our inability to model the entire set of ground truths from the real world

[47] A known limitation of this approach is the 'snowballing effect' of model hallucinations where the model over-commits to its early mistakes, leading to further ones. Zhang et al. (2023) record how ChatGPT and GPT-4 can identify 67% and 87% of their own mistakes, respectively.

[48] "Post-training alignment was applied to reduce hallucination (among other factors) but was also found to reduce calibration" (Kalai et al. 2024 p.3, OpenAI 2024b).

[49] For further exploration of calibration methods, see Wang (2024), which provides a comprehensive review of modern approaches to model calibration.

[50] We cannot view transparency efforts that indicate confidence levels or areas of uncertainty in model responses as sufficient standalone risk management techniques when end users are directly involved. Model developers still deploy tactics of tonality, hedging, and cautionary disclaimers that only exacerbate overreliance (see footnote 44). User perceptions can only be shaped through sincere efforts to avoid such dark patterns in product design (Sandhaus 2023).

**Caveats:** Although calibration metrics can provide valuable insights, their applicability is often limited by deployment conditions that are novel or untested. For example, models trained on general datasets may encounter input distributions that are significantly different from those seen during training, known as distribution shifts. Current methods, however, can still be employed with an understanding of their limitations.[51] For instance, approaches like conformal prediction[52] may be used to guarantee error bounds to uncertainty quantification, but it does not account for distribution shifts. Given these limitations, it therefore becomes crucial to disclose sufficient information on the distribution of training data.

**Transparency about training characteristics can empower downstream deployers to refine uncertainty scores further.** Fine-tuning the model to anticipate and adapt to unique deployment conditions can thus help calibrate model predictions and provide reliable estimates of uncertainty. Where reliable uncertainty measures cannot be established, another appropriate safeguard, such as explicit warning systems and/or model ensembles, should be implemented instead (Rudner and Toner 2024, Gawlikowski et al. 2023).

### Risk Assessment and Likelihood

Modeling risks from misinformation requires not just an estimation of the probability of model outputs containing misinformation, but also the likelihood of generated misinformation leading to undesired impacts.

> **Misinformation Generation vs. Misinformation Impacts**
>
> A model generating misinformation does not necessarily pose misinformation risks on its own. The context of model deployment and model use are often what drive misinformation risks. For example, a model deployed through an API and used to assist in fictional writing would not be susceptible to risks posed by misinformation in model outputs. On the other hand, a model that is used to inform consequential decisions (e.g., legal tasks) or one that is deployed in contexts that may foster the rapid spread of misinformation (e.g., AI-enabled social media platforms), would be highly susceptible to misinformation risks reaching an intolerable level.

---

[51] For instance, calibration metrics such as the Expected Calibration Error (ECE) and Brier scores can help assess model uncertainty. ECE measures how well a model's confidence matches its actual accuracy, while Brier scores evaluate the precision of probabilistic predictions by penalizing both over and underconfidence. Alternatively, deterministic methods can be employed during model training to flag high uncertainty for specific input types, including known adversarial scenarios, to increase robustness.

[52] Conformal prediction, a framework that provides mathematically rigorous uncertainty estimates, works by comparing new inputs to a calibration set of known examples, though it is important to note that its guarantees hold only when the new data follows the same distribution as the training data, i.e., it does not account for distribution shifts.

**Misinformation Generation**

Evaluating a model's misinformation levels can be done by measuring the factual accuracy of generated output, and by performing reasoning and memory hallucinations tests (Gadiko 2024). It is also possible to examine elements of how misinformation may have originated by evaluating the model's tendency to generate misinformation using internal knowledge and external sources, and understanding the model's inclination to hallucinate across different tasks (see Sections 6.1 and 6.2 of Huang et al. 2024). These methods can help identify the model's likelihood of generating misinformation, but for a closer to complete picture of estimated risk, it is necessary to estimate the likelihood of its impact (through sufficient reach).

**Misinformation Impacts**

Evaluating the consequences (or impact) of misinformation is a far more challenging task and requires the consideration of multiple variables. The evaluations of these recommended variables would likely result in estimates that should be transparent in accounting for uncertainty. Appropriate threat models can be designed from these variables to determine the pervasiveness and impact of such frontier AI risks.

- **Severity:** The severity of misinformation can be derived from the context of model use, and can help establish context-based thresholds.

    - Pan et al. 2023 determine one such threat model to determine *misinformation pollution,* which reveals the efficiency of LLMs to act as misinformation generators for malicious use, degrading the performance of Open-domain Question Answering (ODQA)[53] by up to 87% . They extend this to unintentional scenarios of misinformation spread, and demonstrate a 5% to 15% degradation in ODQA performance. While these findings may be specific to ODQA systems, the significant impacts outlined point to its potential to cause a similar scale of harm in related scenarios.

    - The **reversibility** of misinformation impact should also be used as one of the determining factors of misinformation severity. That is, irreversible impacts should generally be considered more severe than reversible impacts, all other things being equal.

- **Spread:** The spread of misinformation can be measured by evaluating the level of misinformation reach. Open AI utilizes the Breakout Scale (Nimmo 2020), which categorizes impact on a six-level scale (e.g., Category 1: spread within one community on one platform, Category 6: triggers a policy response) to assess the impacts of covert influence operations.

---

[53] Open-domain question answering (ODQA) systems must handle queries spanning diverse subjects without predefined context or domain-specific rules. Unlike closed-domain systems, which operate within limited subject areas, ODQA systems need to address factual questions from any field without having the relevant information readily available in their code.

- It is important to consider the likelihood of **acceleration potential** (the speed at which the misinformation spreads) in misinformation spread risk assessments.

- Tracking model use and behaviours by API-level monitoring, especially in tools integrated into an existing platform (e.g., social media) can help anticipate potential misinformation spread (e.g., the number of active users on a platform).[54]

- Additional **impact assessment techniques** could entail the extrapolation of similar analyses measuring misinformation impacts via social media platforms, the extension of prevalent surveys examining user perceptions of and trust in AI, scenario-writing (Kieslich et al. 2024), etc.

- Measuring **trust in media and information** can be used to evaluate the effects of misinformation on information ecosystem pollution (Solaiman et al., 2024).

- Bommasani (2024) demonstrates other similar considerations in delineating different tiers for foundation models to appropriately apply regulations.

**Threshold Recommendations**

> **Proposed Threshold for Misinformation Risks**
>
> Model outputs with misinformation causing immediate severe harms that can be further exacerbated by accelerated spread can be regarded as presenting an intolerable risk.

E.g. Misinformation on vaccines in a pandemic can cause harm to an individual's health and also result in societal harm from contagion, and other secondary harms of economic losses, resource constraints, etc. from a lengthening pandemic.

We recommend using an ensemble of model competence scores and risk estimates evaluated against the potential spread of model use to determine application-specific or domain-specific risk tolerances. We recommend such a domain specific-approach because misinformation in model outputs that recommend instructions to care for house plants might result in harms with lower magnitudes of severity, while similar rates of misinformation errors in model outputs for healthcare recommendations could result in more severe outcomes.

---

[54] In cases where the model and the platform it is integrated into are owned and operated by the same organizations, misinformation spread may be further exacerbated. Proactive measures like automated content evaluation tools must be integrated on these platforms. (Eg. Extrapolating Zhou et al. (2023)'s work to identify AI-generated misinformation) and should be subject to third party oversight. Such scenarios will also benefit from accompanying governance considerations to determine impact. For instance if the platform has over 45 million monthly active users, the EU classifies them in the tier of Very Large Online Platforms or VLOPs and accompanying provisions of the Digital Services Act must follow (EC 2022).

Based on estimates of model uncertainty and risk for an identified deployment context, we recommend setting conservative thresholds (with a low risk tolerance) for high-risk model application areas[55] (e.g., in healthcare diagnostics, or criminal sentencing) where the severity of risk is magnified.[56]

For instance, **for model evaluation results that show high uncertainty and high risk estimates, we recommend that model deployment be disallowed for sensitive applications** where the risk-benefit tradeoff cannot be justified. These levels can be sectorally established; for instance, in the case of a healthcare application, Hjelle (2016) recommends a 95% device reliability score with 95% confidence as a common product validation threshold. Specialized thresholds can be established for specific model deployments informed by previous precedent that reflect the risk appetites peculiar to the sector. [57]

---

[55] Model error rates may be susceptible to domain specific regulatory requirements (e.g., the FDA's medical device premarket approval requirements, FDA 2019 ).

[56] Any such applications (including but not limited to diagnostic systems), irrespective of the levels of certainty or misinformation, must include and empower humans in the loop to make better decisions. But we want to highlight other widespread applications (e.g.,the Google search engine AI overview) which could be doling out incorrect health advice that shouldn't be tolerated (Minsberg 2024). Although recent precedent (Lemley et al. 2023) makes it seem unlikely that such models will be held liable citing their (hidden) product warnings as sufficient cause for deniability, these legislative/regulatory leniencies do not preclude the resulting risks from being intolerable.

[57] Taking the example of the health sector again, specialized risk tolerance levels may look like the FDA's 99.99% reliability requirements with a 95% confidence for emergency-use injectors, which is equivalent to 1/100,000 chances of failure as the requisite safety level. See FDA (2020).

# Conclusion

The definition and operationalization of intolerable risk thresholds for frontier AI models are critical steps in ensuring the safe and responsible development of frontier models through robust governance efforts. This paper provides a comprehensive review of AI risk literature and practical recommendations for stakeholders to participate in the threshold setting exercise. However, it is important to recognize that the field of AI is rapidly evolving, and our understanding of potential risks and appropriate thresholds will need to be continuously updated.

Moving forward, it is crucial for all stakeholders — including industry, governments, academia, and civil society — to collaborate in refining and implementing these thresholds. This will require ongoing research, open dialogue, and a commitment to transparency and accountability in AI development and deployment. As the competition intensifies in the global AI race, the proactive identification and mitigation of intolerable risks will be essential in harnessing the benefits of AI while safeguarding society from potential harms.

CPR (2023) OPWNAI: Cybercriminals Starting to Use ChatGPT. Check Point Research, https://research.checkpoint.com/2023/opwnai-cybercriminals-starting-to-use-chatgpt/

Andrew Critch and Stuart Russell (2023) TASRA: a taxonomy and analysis of societal-scale risks from AI. arXiv, https://arxiv.org/abs/2306.06924

Paulo Rupino Cunha and Jacinto Estima (2023) Navigating the Landscape of AI Ethics and Responsibility. Springer Nature, https://link.springer.com/chapter/10.1007/978-3-031-49008-8_8

Jeffrey Dastin (2018) Amazon scraps secret AI recruiting tool that showed bias against women. Reuters, https://www.reuters.com/article/us-amazon-com-jobs-automation-insight/amazon-scraps-secret-ai-recruiting-tool-that-showed-bias-against-women-idUSKCN1MK08G/

Jiawen Deng, Jiale Cheng, Hao Sun, Zhexin Zhang, and Minlie Huang (2023) Towards Safer Generative Language Models: A Survey on Safety Risks, Evaluations, and Improvements. arXiv, https://arxiv.org/abs/2302.09270

Homayoon Dezfuli, Allan Benjamin, Christopher Everett, Martin Feather, Peter Rutledge, Dev Sen, and Robert Youngblood (2014) NASA System Safety Handbook, Volume 2: System Safety Concepts, Guidelines, and Implementation Examples. National Aeronautics and Space Administration, https://ntrs.nasa.gov/api/citations/20150015500/downloads/20150015500.pdf

DSIT (2023a) Emerging processes for frontier AI. UK Department for Science, Innovation & Technology, https://assets.publishing.service.gov.uk/media/653aabbd80884d000df71bdc/emerging-processes-frontier-ai-safety.pdf

DSIT (2023b) Capabilities and risks from frontier AI. UK Department for Science, Innovation & Technology, https://assets.publishing.service.gov.uk/media/65395abae6c968000daa9b25/frontier-ai-capabilities-risks-report.pdf

DSIT (2024a) Frontier AI Safety Commitments. AI Seoul Summit 2024. UK Department for Science, Innovation & Technology, https://www.gov.uk/government/publications/frontier-ai-safety-commitments-ai-seoul-summit-2024/frontier-ai-safety-commitments-ai-seoul-summit-2024

DSIT (2024b) Seoul Ministerial Statement for advancing AI safety, innovation and inclusivity: AI Seoul Summit 2024. UK Department for Science, Innovation & Technology, https://www.gov.uk/government/publications/seoul-ministerial-statement-for-advancing-ai-safety-innovation-and-inclusivity-ai-seoul-summit-2024/seoul-ministerial-statement-for-advancing-ai-safety-innovation-and-inclusivity-ai-seoul-summit-2024

Abhimanyu Dubey, Abhinav Jauhri, Abhinav Pandey, Abhishek Kadian, Ahmad Al-Dahle, Aiesha Letman, Akhil Mathur, Alan Schelten, Amy Yang, Angela Fan, Anirudh Goyal, Anthony Hartshorn, Aobo Yang, Archi Mitra, Archie Sravankumar, Artem Korenev, Arthur Hinsvark, Arun Rao, Aston Zhang, Aurelien Rodriguez, Austen
54

FDA (2019) Premarket Approval (PMA). US Food and Drug Administration, https://www.fda.gov/medical-devices/premarket-submissions-selecting-and-preparing-correct-submission/premarket-approval-pma

FDA (2020) Technical Considerations for Demonstrating Reliability of Emergency-Use Injectors Submitted under a BLA, NDA or ANDA: Guidance for Industry and Food and Drug Administration Staff - DRAFT GUIDANCE. US Food and Drug Administration, https://www.fda.gov/media/137158/download

Roger Flage, Terje Aven, Enrico Zio, and Piero Baraldi (2014) Concerns, Challenges, and Directions of Development for the Issue of Representing Uncertainty in Risk Assessment. Risk Analysis, https://onlinelibrary.wiley.com/doi/10.1111/risa.12247

FLI (2017) Asilomar AI Principles. Future of Life Institute, https://futureoflife.org/2017/08/11/ai-principles/

Lukas Fluri, Daniel Paleka, and Florian Tramèr (2023) Evaluating superhuman models with consistency checks. arXiv: https://arxiv.org/abs/2306.09983

Aditya Gadiko (2024) Understanding and Addressing AI Hallucinations in Healthcare and Life Sciences. International Journal of Health Sciences, https://carijournals.org/journals/index.php/IJHS/article/view/1862/2238

Jakob Gawlikowski, Cedrique Rovile Njieutcheu Tassi, Mohsin Ali, Jongseok Lee, Matthias Humt, Jianxiang Feng, Anna Kruspe, Rudolph Triebel, Peter Jung, Ribana Roscher, Muhammad Shahzad, Wen Yang, Richard Bamler and Xiao Xiang Zhu (2023) A survey of uncertainty in deep neural networks. Artificial Intelligence Review, https://link.springer.com/article/10.1007/s10462-023-10562-9

Shefali Girish and Meg Avery (2023) Data Cooperative: Enabling Meaningful Collective Negotiation of Data Rights for Communities. SSRN, https://papers.ssrn.com/sol3/papers.cfm?abstract_id=4414473

Josh A Goldstein, Jason Chao, Shelby Grossman, Alex Stamos, and Michael Tomz (2024) How persuasive is AI-generated propaganda? PNAS Nexus, https://academic.oup.com/pnasnexus/article/3/2/pgae034/7610937

Google DeepMind (2024) Introducing the Frontier Safety Framework. Google DeepMind, https://deepmind.google/discover/blog/introducing-the-frontier-safety-framework

Nico Grant (2024) Google's A.I. Search Errors Cause a Furor Online. New York Times, https://www.nytimes.com/2024/05/24/technology/google-ai-overview-search.html

Ryan Greenblatt, Carson Denison, Benjamin Wright, Fabien Roger, Monte MacDiarmid, Sam Marks, Johannes Treutlein, Tim Belonax, Jack Chen, David Duvenaud, Akbir Khan, Julian Michael, Sören Mindermann, Ethan Perez, Linda Petrini, Jonathan Uesato, Jared Kaplan, Buck Shlegeris, Samuel R. Bowman, and Evan Hubinger (2024) Alignment faking in large language models. arXiv, https://arxiv.org/abs/2412.14093
56

Aisha Khatun and Daniel G. Brown (2023) Reliability Check: An Analysis of GPT-3's Response to Sensitive Topics and Prompt Wording. arXiv, https://arxiv.org/pdf/2306.06199

Heidy Khlaaf, Sarah Myers West, and Meredith Whittaker (2024) Mind the Gap: Foundation Models and the Covert Proliferation of Military Intelligence, Surveillance, and Targeting. arXiv, https://arxiv.org/abs/2410.14831

Kimon Kieslich, Nicholas Diakopoulos, and Natali Helberger (2024) Using Scenario-Writing for Identifying and Mitigating Impacts of Generative AI. arXiv, https://arxiv.org/abs/2410.23704

Kevin Klyman (2024) Acceptable Use Policies for Foundation Models. arXiv, https://arxiv.org/abs/2409.09041

Leonie Koessler, Jonas Schuett, and Markus Anderljung (2024) Risk thresholds for frontier AI. arXiv, https://arxiv.org/abs/2406.14713

Leonie Koessler, and Jonas Schuett (2023) Risk assessment at AGI companies: A review of popular risk assessment techniques from other safety-critical industries. arXiv, https://arxiv.org/pdf/2307.08823

Rudolf Laine, Bilal Chughtai, Jan Betley, Kaivalya Hariharan, Jeremy Scheurer, Mikita Balesni, Marius Hobbhahn, Alexander Meinke, Owain Evans (2024) Me, Myself, and AI: The Situational Awareness Dataset (SAD) for LLMs. arXiv, https://arxiv.org/abs/2407.04694

Jeff Larson, Surya Mattu, Lauren Kirchner and Julia Angwin (2016) How We Analyzed the COMPAS Recidivism Algorithm. Propublica, https://www.propublica.org/article/how-we-analyzed-the-compas-recidivism-algorithm

Mark A. Lemley, Peter Henderson, and Tatsunori Hashimoto (2023) Where's the Liability in Harmful AI Speech? SSRN, https://papers.ssrn.com/sol3/papers.cfm?abstract_id=4531029

Nathaniel Li, Alexander Pan, Anjali Gopal, Summer Yue, Daniel Berrios, Alice Gatti, Justin D. Li, Ann-Kathrin Dombrowski, Shashwat Goel, Long Phan, Gabriel Mukobi, Nathan Helm-Burger, Rassin Lababidi, Lennart Justen, Andrew B. Liu, Michael Chen, Isabelle Barrass, Oliver Zhang, Xiaoyuan Zhu, Rishub Tamirisa, Bhrugu Bharathi, Adam Khoja, Ariel Herbert-Voss, Cort B. Breuer, Andy Zou, Mantas Mazeika, Zifan Wang, Palash Oswal, Weiran Liu, Adam A. Hunt, Justin Tienken-Harder, Kevin Y. Shih, Kemper Talley, John Guan, Russell Kaplan, Ian Steneker, David Campbell, Brad Jokubaitis, Alex Levinson, Jean Wang, William Qian, Kallol Krishna Karmakar, Steven Basart, Stephen Fitz, Mindy Levine, Ponnurangam Kumaraguru, Uday Tupakula, Vijay Varadharajan, Yan Shoshitaishvili, Jimmy Ba, Kevin M. Esvelt, Alexandr Wang, and Dan Hendrycks (2024a) The WMDP Benchmark: Measuring and Reducing Malicious Use With Unlearning. arXiv, https:/arxiv.org/abs/2403.03218

Nathaniel Li, Alexander Pan, Anjali Gopal, Summer Yue, Daniel Berrios, Alice Gatti, Justin D. Li, Ann-Kathrin Dombrowski, Shashwat Goel, Long Phan, Gabriel Mukobi, Nathan Helm-Burger, Rassin Lababidi, Lennart Justen, Andrew B. Liu, Michael Chen, Isabelle Barrass, Oliver Zhang, Xiaoyuan Zhu, Rishub Tamirisa, Bhrugu
59

# Appendix A

## Background on Risk Categorizations

**In National Policies:**

**US policy** introduces **"unacceptable negative risk levels"** in NIST's AI Risk Management Framework (AI RMF), which defines the highest risk tier prompting immediate cessation in model development and deployment. The NIST AI RMF states, "In cases where an AI system presents unacceptable risk levels – such as where negative impacts are imminent, severe harms are actually occurring, or catastrophic risks are present – development and deployment should cease in a safe manner until risks can be sufficiently managed" (NIST 2023, p. 8).

The analogous nomenclature from the **EU AI Act** characterizes such risk outcomes as **systemic risks** from high-impact capabilities of general-purpose AI models, having a significant impact on the Union market due to their reach (see, Articles 5 and 52 in EP 2024).

The **UK**'s International Scientific Report on the Safety of Advanced AI (Bengio et al. 2024, 2025), however, introduces a helpful classification of AI risks into three categories: technical risks, misuse risks, and societal risks. These categories can be summarized as follows: **technical risks**, which include system limitations and malfunctions like bias and discrimination, confabulation, or design failures; **misuse risks,** which involve malicious activities such as CBRN proliferation or manipulation of public opinion; and finally, **societal risks,** that encompass broader impacts, including IP and copyright issues, environmental and socioeconomic harms, and privacy violations, with some risks cutting across these categories.

*All three documents list similar risk outcomes, albeit with distinct terminology.*

**From Industry Policies: Misuse Risks and Safety Levels**

Organizational risk management policies stratify capability-based risks into various tiers. In our interpretation of risk types and impacts, we see the intolerable risk level emerge well before the highest tier/level of risks that they measure against.

1. In its **Responsible Scaling Policy (RSP)**, **Anthropic** categorizes capability thresholds as **"red line" and "yellow line" capabilities** (Anthropic n.d.), with corresponding AI Safety Level (ASL) standards. Red line capabilities refer to anticipated model abilities that may appear in future versions of the model and would present too much risk if deployed under current ASL-2 safety measures. Anthropic has committed to developing a new set of ASL-3 safety measures to sufficiently manage and mitigate models with red-line capabilities. Anthropic has also defined qualitative capability thresholds for specific model capabilities (CBRN weapons, autonomous AI research and development, and cyber operations). For example, the CBRN weapon threshold is defined as "the ability to significantly help individuals or groups with basic technical backgrounds (e.g., undergraduate STEM degrees) create/obtain and deploy CBRN weapons" (Anthropic 2024 p.3). According to the company's internal policies, such capabilities would warrant the application of Level 3 ASL or ASL-3 to prevent model misuse (Anthropic 2024). These thresholds are evaluated by first conducting a preliminary assessment to determine if a model is "notably more capable" than the latest model that has been comprehensively tested. Models that are effectively more capable (4x or more in Effective Compute or six months worth of fine-tuning) undergo a comprehensive assessment containing threat model mapping for each capability threshold, empirical tests for capability evaluation, elicitation testing without safety mechanisms, and likelihood forecasting (Anthropic 2024). Anthropic maintains a dynamic risk scorecard that reflects pre- and post-mitigation evaluation results for each of the tracked capability categories. The risk levels inform Anthropic's decision to enforce certain safety baseline actions based on pre- or post-mitigation risk scores. For example, models can only be deployed if they are determined to have a post-mitigation risk score of "medium" or below.

2. The **OpenAI preparedness framework** categorizes thresholds using a **qualitative scale (low, medium, high, and critical)** with definitions for each of their four tracked capability categories (cybersecurity, persuasion, CBRN, and model autonomy). The "high" and "critical" categories here could be treated as analogous to intolerable levels of threat in model capabilities that require heightened risk mitigation efforts or an altogether pause in development (OpenAI 2023b). For example, a model that is considered to have a "high" level of cybersecurity capability risk is defined as a "tool-augmented model (that) can identify and develop proofs-of-concept for high-value exploits against hardened targets without human intervention, potentially involving novel exploitation techniques, OR provided with a detailed strategy, the model can end-to-end execute cyber operations involving the above tasks without human intervention" (OpenAI 2023b p.8). A model's overall capability score is determined by the highest score in any of the tracked categories.

3. Google DeepMind's Frontier Safety Framework (Google DeepMind 2024) outlines model **critical capability levels (CCLs)**, which are identified with preliminary model evaluations for various risk domains (autonomy, biosecurity, cybersecurity, and machine learning R&D). Each risk domain CCL is described and includes the rationale behind the categorization. For example, "bio expert enablement level 1" is described as "capable of significantly enabling an expert (i.e., PhD or above) to develop novel biothreats that could result in an incident of high severity" (Google DeepMind 2024 p.5). The safety evaluations based on this framework help the organization to decide on security and/or deployment mitigations to address severe risks posed by their models and pause deployment/development if the evaluated model capability supersedes the institution of necessary risk mitigation levels (Google DeepMind 2024). The model developer states, "The Framework is exploratory and based on preliminary research. We expect it to evolve substantially as our understanding of the risks and benefits of frontier models improves, and we will publish substantive revisions as appropriate" (Google DeepMind 2024 p.6).

Across the three examples described above, there is relative consensus as to the key risk categories that are considered in determining capability thresholds. All three discuss the general risk categories of CBRN weapons, cyber operations, and model autonomy. The OpenAI preparedness framework additionally considers *persuasion,* and notes that their framework includes *deception* and social engineering evaluations as part of the persuasion risk category. Anthropic's Responsible Scaling Policy includes the footnote, "We recognize the potential risks of highly persuasive AI models. While we are actively consulting experts, we believe this capability is not yet sufficiently understood to include in our current commitments."

### Taxonomic Fragmentation

The heterogeneous ways in which AI risks are characterized across different geographies, application domains, academic disciplines, and taxonomies appeared as a significant challenge when we began this project.

We began our research on intolerable risks in an attempt to rectify this incoherence in the classification of AI harms. This included efforts to both compile existing taxonomies in an exhaustive manner, as well as efforts to pick from other risk databases to help define the scope of risks to which our methodology to establish intolerable risk thresholds could be applied (Slattery et al. 2024, Zeng et al. 2024, Vidgen et al. 2024; Weidinger et al. 2022). However, over the course of these efforts to harmonize different risk categories, it became increasingly apparent that the distinct taxonomies were not necessarily a shortcoming but an opportunity to reconsider the limitations of a centralized framework to approach this problem.

Therefore, we abandoned the search for an elusive universal taxonomy and instead adopted a principled approach toward operationalizing intolerable risk thresholds, with an initial set of risk categories that we determined to be in scope. We acknowledge the immense benefits of a standardized taxonomy that could enable better uniformity and comparability in the risk assessment and reporting exercise for better AI governance. However, we also advocate for such a taxonomy to be informed not just by literature but also be reflective of the national, cultural, and organizational appetites and attitudes towards risk.

### Risk Entanglements and Subjectivities

To substantiate our claim about the utility and value of these distinct, sometimes even divergent, characterizations of risk, we lean on the unique perspective gathered from applying relational risk theory to map risk entanglement. In this work, von Scheve and Lange's (2023) track how different **risk actors** (be they individuals, institutions, or geographies) may characterize the same risk scenario in entirely different ways. They borrow the concepts of **risk objects** and **objects at risk** (Boholm and Corvellec, 2011) to demonstrate the social and cultural paradigms that shape our understanding of risk.

Applying this to the risks from frontier AI models, the **risk objects** would point to the various sources of risk, such as model capabilities (autonomy, persuasion, tool access), model limitations/failures (bias, loss of control), deployment context (number of users, domain vulnerability), and other model propensities. **Objects of risk**, on the other hand, refer to the impact on valued artifacts, which in AI risks could take the form of loss of human lives, damage to property and infrastructure, socioeconomic losses, etc. Finally, **risk actors** could be identified as frontier model developers, downstream deployers, government actors, end-users, etc.

For instance, developers may perceive the risk object of model hallucinations as a hindrance to model adoption or brand perception, while downstream actors may find the object of risk as the undisclosed nature of training data that limits their ability to reliably fine-tune the model to produce robust results. The object of risk for an end-user however, might no longer be related directly to the model performance and instead could be more behavioral, like their overreliance on model outputs. For media institutions however, the same risk may present a more macro harm in the form of an irreparable poisoning of the information landscape or the loss of trust in social institutions. Therefore, we believe that accounting for the subjectivity in risk characterization for different risk actors, and sources of risk (risk objects), should be central to the risk estimation exercise.

Building on the relational risk framework, von Scheve and Lange (2023) additionally demonstrate the critical process of **risk entanglement** arising from the unique relationships between different risk actors. Commenting on how risk actors relate to each other in a network, they characterize the impacts of the framing of risks:

> "For decision makers, a potential harm appears as a risk, whereas for those who are affected by the decision, this harm appears as a danger."

For instance, AI safety frameworks from industry stratify capability-based risks into various tiers that reflect the developers' appetites for risk. In our interpretation of risk types and impacts, however, we see that intolerable risk levels emerge well before the highest tiers or levels of risks that these industry frameworks establish. We suspect that governmental agencies and regulators may similarly have divergent views on this stratification. We therefore detail our submission to determining intolerable risk thresholds based on this dynamic and subjective interpretation of risk without situating them in any one taxonomy, as explained further in Section 2.3 and 3.

# Appendix B

## Types of Thresholds

### Compute Thresholds

Compute thresholds are often measured with floating-point operations per second (FLOPS). Compute thresholds for models when set - (i) **above the frontier** (currently $10^{26}$ FLOPs) would include novel capabilities that are difficult to predict, (ii) **at the frontier** (currently $10^{25}$ FLOPs) may already include dangerous capabilities, (iii) **below the frontier** (currently $10^{24}$ FLOPs) would be the most cautious approach, but may create unnecessary regulatory burdens (Heim and Koessler 2024). However, there is also a lack of clarity in the approaches used to measure FLOPs for various types of systems (C4AI 2024). Compute thresholds may be most useful as an *initial* metric to identify models that require further regulatory oversight and evaluation, but there are many factors beyond compute that contribute to model capability. Enhancing data quality, implementing model optimization techniques, and utilizing novel model architectures can lead to increased model capabilities without requiring an increase in model compute (C4AI 2024). Algorithmic efficiency improvements can also lead to a decrease in the levels of compute required for certain model capabilities, which can be measured by the metric "effective compute" (Heim and Koessler 2024).

### Capability Thresholds

Capability thresholds are the most commonly adopted approach used in AI frontier safety frameworks to identify the different risks from specific capabilities and stratify them into different levels based on the potential magnitude of impact (as summarized in Appendix A). Because of the dual-use nature of foundation models, risk often stems from capability (Koessler et al. 2024). Advanced models may amplify societal risks if they are exploited to increase the effective ability of malicious actors to execute attacks, or are deployed to autonomously execute attacks (e.g., cyber and CBRN attacks) (Barrett et al. 2024b, UK AISI 2024).

### Risk Thresholds

Risk is often defined in terms of likelihood (i.e., the probability of an event), and severity of harm (i.e., the magnitude of impact). Comprehensive risk models containing all possible risk scenarios are extremely difficult to develop, and it is recommended to start with a limited and defined number of risk scenarios (Koessler et al. 2024). Regardless of the risk measurement method, it is important that risk thresholds are operationalized and specific enough such that multiple evaluators with access to the same resources would agree on the risk threshold determination of an evaluated model (DSIT 2023a). A common proxy used by model providers to measure risk is "model capability" (see Anthropic 2024, Google DeepMind 2024).

Another common proxy measure for risk is "compute power," defined in both Executive Order 14110[58] and the EU AI Act.

---

[58] The majority of this report was drafted and finalized prior to the rescission of Executive Order 14110 on January 20, 2025.

# Appendix C

## Best Practices in Model Evaluations

Boyarskaya et al. (2020) recommend the adoption of dimensions for responsible innovation proposed by Stilgoe et al. (2013) to better estimate harmful impacts from AI systems. These dimensions of *anticipation, reflexivity, inclusion*, and *responsiveness* would be essential features of any robust risk assessment and management framework.

1. **Anticipating Risks**

**Designing the Right Evaluations:**

In accepting current evaluation practices while determining thresholds, there is an implicit assumption of the general reliability of benchmarks. However, model benchmarking and evaluation practices are still nascent and sometimes provide insufficient or incorrect estimations of model capabilities. This should be improved by:

- **Ensure Construct Validity:** It is important to ground model evaluations in use cases and design evaluation benchmarks that are curated for specific domains, and designed with input from the community. (For examples in practice in legal and medical fields, see Guha et al 2023, Nayak et al. 2023).

- **Involve End-users:** Citing prompt sensitivity as a crucial challenge, Kapoor, Henderson et al. (2024) recommend the involvement of end users in the evaluation process to adequately determine model capabilities.

**Considerations for Robust Evaluations:**

- **Map harms throughout product life cycle:** Model risk is accrued at every step of the AI development pipeline, from the choice of data labels to the domain of their deployment. It is therefore necessary to identify and document the risks at every stage of development to inform developer decision-making at each point. The choice of using sensitive personal data in frontier models that could be deployed in surveillance technologies must be an explicitly stated choice to ensure it falls in line with organizational appetite for risk.

2. **Reflexivity**

Designing sound risk anticipation frameworks for foundational technologies that operate at several layers of abstraction or distance from users is ultimately a design challenge. There is a constant need to reassess assumptions, state uncertainties, and leave room for *unknown unknowns* in how we model the trajectory of these frontier technologies. For instance, while most forecasting studies take into account the growing access to compute capabilities, the decreasing costs of model inference, the anticipated pace of workplace automation, etc., the compounding impact of these distinct effects is not entirely discernable, and therefore deployment decisions need to register the sum of externalities to inform every stage of release.

- **Map Harms to Known Algorithms:** Language is inherently political and its inherent choices of representation hold power over society, thereby bestowing a peculiarity to LLMs that is unlike other transformative technologies. Naturally the creation of language models from text corpora will stand to mimic the same biases that exist in society. This is a known harm of representation/discrimination/bias from NLP and LLM technologies, and its amplification within AI models is not surprising (Caliskan 2017, Schwartz et al. 2019). Therefore, applying language technologies to domains of high sensitivity, such as psychology or journalism, only exacerbates the manifestation of these inherent biases in the absence of due safeguards and fine-tuning. Therefore, applications of algorithms with known harms in high-impact domains need to be evaluated against stricter thresholds, as opposed to, say, their application in e-commerce recommendations.

    - **Decentralizing decision-making** and sufficiently adapting design by including end users and affected stakeholders before choosing to deploy these models would be crucial in this juncture. However, participation needs to be preceded by appropriately familiarizing these groups to known harms to elicit their informed perspective.

    - **Acceptable Use Policies (AUP):** While somewhat antithetical to decentralizing powers, AUPs could be a necessary short-term solution, especially in open foundational models, until we design better oversight mechanisms. Klyman (2024) indexes some such vulnerabilities by analyzing AUPs from foundational model developers that place restrictions on their use towards producing misinformation, generating misinformation, or restricting the scale of model deployments, or their application in highly regulated industries.

3. Inclusion

- **Beyond Technical Assessments:**

    While the "generality" of frontier models allow for wide applicability, these base models are not bereft of values, and neutral technical assessments would therefore be insufficient to evaluate their readiness for deployment. While certain types of risks are more amenable to quantitative thresholds along vectors such as accuracy, probability, expertise etc., other risks, especially systemic risks with longer timelines, require qualitative measures to assess their threat levels. Across risk categories however, there must be avenues for assessments to be defined and delivered by developers, experts, and impacted communities as appropriate.

    - **Resisting a tech-first approach:** As Kapoor, Bommasani et al. (2024) recommend in their marginal risk assessment framework, it is necessary to establish the risk of identified harms for different populations or domains in the absence of frontier model applications. For instance, when training models on data collected from vulnerable populations,

- developers should ask, what are the risks of breaches to information integrity, and how would they be different from data safety without AI application?
  - **Multi-disciplinary approaches to assessment:** In order to reliably anticipate model interactions in the contexts of their deployment and center *all* end users, there is a need for cross-disciplinary approaches to AI evaluation (Schwartz et al. 2022). Widespread application of sociotechnical modalities are an encouraging sign towards such an approach; however, such applications need to permeate into industry best practices more effectively to improve risk anticipation. Apart from the application of social science frameworks, transferring learnings from the fields of cybersecurity, finance, healthcare, etc. will also be essential in this mapping exercise. As Khlaaf et al. (2024) postulate, it would be necessary to treat models trained on sensitive data with the same restrictions as those placed on the data itself.

- **Beyond AI Safety**

  Community-centric frameworks and participatory approaches to AI evaluation are gaining steady reception despite the known challenges in implementing them at scale. However, the ethos of inclusive design needs to be operationalized at earlier stages of model development. For instance, Mohamed et al. (2020) recommend the application of a decolonial lens to the development cycle by canvassing for the adoption of a critical technical practice, or tapping into the emergence of more specialized community groups with the technical know-how to empower their communities' needs (e.g., Queer in AI, Indigenous voices in AI, Data Cooperatives. See, Ovalle et al. 2023, Canavera 2023, Girish and Avery 2023 .).

  Inclusion also needs to extend to fast-digitizing, data-rich countries of the global majority that are facing similar challenges to safeguard populations who are relatively recent adopters of digital technologies. Apart from the large-scale impacts this imbalance might have on a country's participation in the global digital economy, it is also the most vulnerable groups that often find themselves to be the earliest subjects in model deployment trials, with no efforts to ensure adequate awareness or require informed consent. There is an urgent need for global majority voices to be given a platform in international forums to advocate for marginalized populations that will be adversely affected (Png 2022, Qadri 2023).

4. **Responsiveness**

In Barrett et al (2025), we present extensive guidance on risk management and mitigation as an accompaniment to the Risk Management Framework from NIST (NIST 2023). Effective risk management requires developer's to coordinate with multiple stakeholders and engage in long-term planning and resourcing to mitigate intolerable AI risks. (For further guidance on responsiveness, see FLI 2017, Srikumar et al. 2024.) Simultaneous resource and risk planning and allocation need to happen through state actors to curb the scale of impact.

- **Ensure that model releases are preceded by a comprehensive risk mitigation strategy** that relies on multiple independent, overlapping safety mechanisms to withstand failures in any particular defence techniques (Bengio et al. 2024, 2025).

- **Institute AI Security Practices:** Robust information security practices need to be established to protect product vulnerabilities. These include but are not limited to:

    - **Adversarial Testing:** Conduct red teaming and adversarial testing on frontier models to identify security risks (e.g., backdoors, AI trojans, or prompt injection attacks), especially for models trained on potentially poisoned public data.

    - **Secure Model Weights:** Recent research from RAND (Nevo et al. 2024) presents detailed recommendations for securing frontier model weights across five defined security levels (SLs). For further guidance, see Measure 1.1 in Barrett et al. (2025).

    - **Privacy and Data Protection:** Identify memorization and ensure contextual integrity when using sensitive data or PII to train models. See Section 3.5 of Solaiman et al. (2024).

- **Maintain Transparency:** As mentioned in the overarching comments in Section 3.1., documented risks and decisions should also be reported transparently to concerned stakeholders.

    - **Communicate Limitations:** Having documented extensively all known risks from the product, it is necessary to paint an honest image of model capabilities so as to encourage end user deliberation on model outputs. This could range from adopting bright patterns in product design (Sandhaus 2023) or recommending extensive responsible practices and acceptable use policies for downstream deployment.

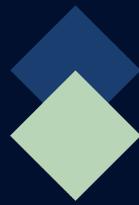